\documentclass{emulateapj}
\usepackage{float,epsfig,psfig}

\newcommand{\LA}{\mbox{\raisebox{-0.6ex}{$\stackrel{\textstyle<}{\sim}$}}}
\newcommand{\GA}{\mbox{\raisebox{-0.6ex}{$\stackrel{\textstyle>}{\sim}$}}}
\newcommand{\cxo}{{\sl Chandra}}

\newcommand{\cha}{{\sl Chandra}}
\newcommand{\ein}{{\sl Einstein}}

\newcommand{\msun}{M$_{\odot}$}
\newcommand{\ergl}{ergs~s$^{-1}$}

\newcommand{\ergcms}{ergs~cm$^{-2}$~s$^{-1}$}

\newcommand{\mdot}{$\dot{M}$}
\newcommand{\ros}{{\sl ROSAT}}
\newcommand{\asca}{{\sl ASCA}}
\newcommand{\elip}{{$D_{25}$}}
\newcommand{\lfir}{$L_{\rm FIR}$}
\newcommand{\lx}{$L_{\rm X}$}
\newcommand{\lb}{$L_{\rm B}$}
\newcommand{\lognlogs}{log($N(>S)$)~--~log($S$)}
\newcommand{\etal}{et al.}

\begin{document}

\title{
The Ultra-Luminous X-ray Source Population
from the Chandra Archive of Galaxies}

\author{
Douglas~A.~Swartz\altaffilmark{1},
Kajal~K.~Ghosh\altaffilmark{1},
Allyn~F.~Tennant\altaffilmark{2},
Kinwah~Wu\altaffilmark{3}
}
\altaffiltext{1}{Universities Space Research Association,
NASA Marshall Space Flight Center, SD50, Huntsville, AL, USA}
\altaffiltext{2}{Space Science Department,
NASA Marshall Space Flight Center, SD50, Huntsville, AL, USA}
\altaffiltext{3}{MSSL, University College London, Holmbury St. Mary, Surrey,
RH5 6NT, UK}

\begin{abstract}
One hundred fifty-four discrete non-nuclear Ultra-Luminous X-ray (ULX) sources,
 with spectroscopically-determined intrinsic X-ray luminosities
 $>10^{39}$~\ergl, are identified in 82 galaxies observed
 with {\sl Chandra}'s Advanced CCD Imaging Spectrometer.
Source positions, X-ray luminosities, and spectral and timing
 characteristics are tabulated.
Statistical comparisons between these X-ray properties and those of the weaker
 discrete sources in the same fields
 (mainly neutron star and stellar-mass black hole binaries)
 are made.
Sources above $\sim$$10^{38}$~\ergl\
 display similar spatial, spectral,
 color, and variability distributions.
In particular, there is no compelling evidence in the sample for
 a new and distinct class of X-ray object such as the
 intermediate-mass black holes.
83\% of ULX candidates have spectra that can be described as
 absorbed power laws with index $\langle \Gamma \rangle = 1.74$ and
 column density $\langle N_H \rangle = 2.24\times 10^{21}$~cm$^{-2}$,
 or $\sim$5 times the average Galactic column.
About 20\% of the ULXs have much steeper indices indicative of a soft,
 and likely thermal, spectrum.
The locations of ULXs in their host galaxies are
 strongly peaked towards their galaxy centers.
The deprojected radial distribution of the ULX candidates is
 somewhat steeper than an exponential disk, indistinguishable from that of
 the weaker sources.
About 5--15\% of ULX candidates are variable during the \cxo\ observations
 (which average 39.5~ks).
Comparison of the cumulative X-ray luminosity functions of the ULXs to
 \cxo\ Deep Field results suggests $\sim$25\% of the sources may be background
 objects including 14\% of the ULX candidates in the sample of spiral galaxies
 and 44\% of those in elliptical galaxies
 implying the elliptical galaxy ULX population is severely compromised by
 background active galactic nuclei.
Correlations with host galaxy properties confirm the number and total X-ray
 luminosity of the ULXs are associated with
 recent star formation
  and with galaxy merging and interactions.
The preponderance of ULXs
 in star-forming galaxies as well as their
 similarities to less-luminous sources suggest they
 originate in a young but short-lived population such as the high-mass X-ray
 binaries with a smaller contribution (based on spectral slope)
 from recent supernovae.
The number of ULXs in elliptical galaxies
 scales with host galaxy mass and
 can be explained most simply as
 the high-luminosity end of the low-mass X-ray binary population.
\end{abstract}


\section{Introduction}

Among the most intriguing objects in the X-ray sky are the discrete non-nuclear
 Ultra-Luminous X-ray sources (ULXs) in nearby galaxies.
This name describes sources considerably more luminous than expected for
 a spherically-accreting object of typical neutron star mass.
Here, ULXs are defined to be those with apparent (i.e., assumed
 isotropically emitting) intrinsic luminosities in excess of $10^{39}$~\ergl\
 in the 0.5--8.0 keV bandpass.

\ein\ X-ray images of galaxies revealed 16 ULX candidates
 (Long \& Van~Speybroeck 1983; Helfand 1984; Fabbiano 1989).
\ros\ has extended this list to nearly 100 (Roberts \& Warwick 2000;
Colbert \& Ptak 2002).
Yet these objects are rare, reported at a rate of only 1--2 per
 galaxy from pointed X-ray observations
 (Colbert \etal\ 2004; Kilgard \etal\ 2002; Colbert \& Ptak 2002;
 Foschini \etal\ 2002; Humphrey \etal\ 2003; Irwin, Athey, \& Bregman 2003)
 and perhaps occurring much less frequently in the nearby Universe as a
 whole (Ptak \& Colbert 2004).
ULXs may represent the high-luminosity end of a continuous distribution
 of typical X-ray sources such as supernovae and X-ray binaries
 (e.g., Grimm, Glifanov, \& Sunyaev 2003) or they
 may include new classes of objects including intermediate mass black holes
 (Colbert \& Mushotzky 1999; Makishima \etal\ 2000; Colbert \& Ptak 2002;
 van~der~Marel 2004), beamed sources (King \etal\ 2001; Georganopoulos \etal\
 2002; K\"{o}rding, Falcke, \& Markoff 2002), and hypernovae (Wang 1999).

Part of the reason ULXs are, as a class,
 poorly understood is that past X-ray observatories lack the combination of high
 angular resolution and moderate spectral resolution needed
 to adequately characterize discrete X-ray sources in nearby galaxies.
The highest spatial resolution instruments aboard both \ein\ and \ros\ provide
 little or no spectral information and their spectroscopic instruments
 have poor angular resolution.
The \asca\ satellite provides moderate spectral resolution but very poor
 angular resolution.
As a result, ULXs are difficult to resolve from other nearby X-ray sources and from
 diffuse emission in the field;
their positions and morphologies are poorly known thus impeding follow-up
 investigations in other wavebands; and ULX X-ray spectra and timing
 properties are often sparsely sampled.

We have undertaken an extensive X-ray spectrophotometric survey of the
 discrete source populations in nearby galaxy fields anchored
 upon archival data
 obtained with the \cxo\ X-ray Observatory Advanced CCD Imaging Spectrometer
 (ACIS).
The main goal of the survey is to systematically search for and
 evaluate the physical properties of ULXs.
The survey encompasses a sample of ULXs large
 enough for meaningful
 statistical analyses; the survey is based upon high spatial resolution
 images that provide accurate celestial positions as well as
 medium-resolution broad-band spectral signatures and time sampling.
The galaxies included in the survey span the range of Hubble morphological types
 and include galaxies of various mass, gas content, dynamical state,
 and evolutionary history; thus
 allowing correlations between ULXs and their local
 environments to be studied.

The galaxy selection process and the methods of X-ray data reduction
 and analysis are described in \S \ref{s:METHODS}.
The resulting sample of galaxies, their properties,
 and the properties of the ULX candidates are given in \S \ref{s:RESULTS}.
The implications of the current survey
 for contemporary physical models for the ULX phenomena is discussed in
 \S \ref{s:DISCUSSION}.

\section{METHODS} \label{s:METHODS}
\subsection{Galaxy Selection} \label{s:METHODS_galSelect}

\cxo/ACIS, operating in imaging mode,
 provides moderate spectral resolution and throughput at the highest
 spatial resolution available in an X-ray telescope facility.
As our primary goals are to obtain accurate celestial positions and to
 perform spectroscopic analysis of ULX candidates,
 ACIS is the ideal instrument for the present study.
To achieve these goals requires a minimum of
 $\sim$50 X-ray counts per source distributed over several spectral
 energy bins.
Thus, our sample includes nearly all available non-grating,
 timed exposure mode, ACIS galaxy observations
 (those listing a galaxy or a supernova as a target in the \cxo\ approved target
 lists\footnote{http://asc.harvard.edu/target\_lists}) publicly-available
 c.~May 2003 with \GA 50 counts expected from
 sources with intrinsic 0.5--8.0 keV luminosities of \lx $=$$10^{39}$~\ergl.
NGC~221, NGC~224, and the Magellanic Clouds are the only galaxies
 purposely excluded from the sample.
These galaxies are known to contain no
 ULXs and require multiple pointings for full coverage (except for the
 dwarf elliptical NGC 221).
There were no non-targeted galaxies in the \cxo\ fields that
 met the selection criterion.

For each candidate galaxy in the archive, the number of counts
 expected from a ULX was estimated
 using the Portable Interactive Multi-Mission Simulator (PIMMS, Mukai 1993);
 the distance, $D$, to the galaxy; the duration, $t$, of the observation;
 and assuming an absorbed power law spectral shape with spectral
 index $\Gamma=1.6$ and with a hydrogen column density, $N_H$, equal to the
 Galactic column density along the line of sight.
The ACIS count rate for this spectral shape is $\sim$$10^{-4}$~c~s$^{-1}$
 for a source flux of $10^{-15}$~\ergcms\ (for a source imaged on the
 back-side illuminated CCDs in the 0.5--8.0~keV band and for
 a typical $N_H/10^{20} \sim 1$--10~cm$^{-2}$.
The count rate is about 30\% lower if the source is imaged on the
 front-illuminated CCDs.)
Thus, the requirement of 50~c from a source with luminosity \lx\
 $=10^{39}$~\ergl\ can be expressed as $t/D^2$\GA 60 where
 $D$ is measured in Mpc.
Typical ACIS observations are $\sim$50~ks duration; limiting the distances to
 host galaxies to \LA 30~Mpc (\LA 13~Mpc for a 10~ks observation).

Distances to the host galaxies were obtained from the literature.
In order of preference, distances are based on the Cepheid period-luminosity
 relation ($\sim$5\% distance uncertainty; see Ferrarese \etal\ 2000),
 $I$-band surface brightness fluctuations (SBF, 10\% uncertainty), on the
 tip of the red giant branch (10\%),
 or on planetary nebula luminosity functions (10\%).
Distances based on globular cluster luminosity functions and
 $K^{\prime}$ and $K_s$-band SBF methods are somewhat more uncertain
 (Ferrarese \etal\ 2000).
In a few cases, distances were based on tertiary distance indicators
 such as the brightest blue stars (see de~Vaucouleurs 1978;
 Karachentsev  \& Tikhonov 1994 for definitions)
 and only recent, well-calibrated, distances were deemed reliable.
Finally, distances tabulated in the catalogue of Tully (1988) were
 used when no other estimate was available.
This catalogue uses
 the Virgo infall model of Tully \& Shaya (1984),
 a variant of the Tully-Fisher relation, that assumes an
 infall velocity of 300 km~s$^{-1}$ for the Local Group,
 $H_o = 75$~km~s$^{-1}$~Mpc$^{-1}$, and
 a Virgo distance of 16.8 Mpc. Distances in the Tully catalogue are most
 reliable for galaxies beyond Virgo.

The duration of the observations, $t$, were taken from the good-time intervals
 provided as part of the X-ray event lists.
Galactic hydrogen column densities along the line-of-sight to the host
 galaxies were taken from  the  HI  map of Dickey \& Lockman (1990) using
 the FTOOL utility {\tt nh} available from HEASARC.

\subsection{Basic Galaxy Properties} \label{s:METHODS_galProps}

Several global properties of the host galaxies were obtained to aid in data reduction
 and for subsequent analysis.
Galaxy morphological type, major isophotal angular diameter
 ($\equiv$\elip, measured at surface brightness level
 25.0 mag sec$^{-2}$ in blue light), major-to-minor isophotal diameter ratio,
 position angle of the major axis, galactic latitude, total (asymptotic)
 $B$-band magnitudes corrected for extinction and redshift, and $(B-V)$
 colors are all taken from the Third Reference Catalogue of Bright Galaxies
 (de~Vaucouleurs~\etal\ 1991, hereafter RC3).
The absolute $B$ magnitudes, ${\rm M}_{\rm B}$, were derived from the
 apparent magnitudes and the
 adopted distances and converted to luminosities, $L_B$, using
 the standard photometric quantities cited in Zombeck (1990).
Galaxy center coordinates were taken from the NASA/IPAC Extragalactic Database
 (NED) and are often based on accurate radio measurements.
Typically, these values differ by no more
 than a few seconds of arc from those reported in RC3.

\subsubsection{Far-Infrared Luminosities}

Far-infrared
 measurements of the host galaxies made by the {\sl Infrared Astronomical
 Satellite} ({\sl IRAS}) were used to derive the far-infrared luminosities of
 galaxies in the sample.
The total flux between 42.4 and 122.5~$\mu$m is approximated by
 $1.26\times10^{-11}(2.58S_{60}+S_{100})$~\ergcms\
 where $S_{60}$ and $S_{100}$ are
 the total flux densities at the 60 and 100~$\mu$m bands, respectively
 (Rice \etal\ 1988).
Flux densities were taken, in order of preference, from the tabulations in
 Ho, Filippenko, \& Sargent (1997), Rice \etal\ (1988), Knapp \etal\ (1989),
 or from NED.

\subsubsection{Nearest Neighbor Distances}

The separation between each host galaxy and its nearest neighbor was taken
 from the work of Ho, Filippenko, \& Sargent (1997) or directly from NED.
We define nearest neighbor to be the closest galaxy on the sky with a
 small relative velocity between the pair ($\Delta v$\LA 500~km~s$^{-1}$).
The projected angular nearest-neighbor separation, $\theta_P$, quoted in
 this work is given in units of the \elip\ isophotal angular diameter.

\subsection{X-ray Data Reduction} \label{s:METHODS_xrayData}

For each X-ray observation,
 events within the \elip\ isophote were extracted from Level~2 event lists
 for analysis.
For events on CCD S4, the {\tt destreak} algorithm\footnote{
 http://asc.harvard.edu/ciao2.1/downloads/scripts/destreak.ps}
 was applied to remove charge randomly deposited along
 pixel rows during readout and data from all CCDs
 were cleaned of bad pixels and columns.
X-ray sources were then located and
 source spectra and time series for events within an $\sim$90\%
 encircled energy radius were extracted along with a background spectrum in
 a surrounding annulus for all sources with estimated luminosities exceeding
 $\sim$$10^{38}$~\ergl.
(These estimates were based on the number of detected counts
 and assume the same spectral shapes as was used for galaxy
 selection, \S\ref{s:METHODS_galSelect}). Some
details of the source finding algorithm are given in Swartz \etal\ (2003).
The source and background region image, spectrum, and light curve were then visually inspected
 for anomolies such as background flares or other sources overlapping the
 extraction regions and appropriate adjustments made.
Sources located within 5\arcsec\ of the
 host galaxy nucleus are not further considered in this work.
X-ray source positions were mapped to second-generation Digitized Sky Survey
 blue (roughly 400--550 nm)
 images of the galaxies in order to identify and reject bright
 (above $\sim$20~mag) foreground stars based on spatial coincidence.
Further analysis of non-X-ray data is in progress and will be
 reported elsewhere.
In the few cases where more than one compatible observation of a galaxy was
 available, datasets were merged using FTOOL utilities to increase
 signal-to-noise.

\subsubsection{Spectral Analysis}

The XSPEC (v.11.2) spectral-fitting package was used for analysis of
 events in PI channels corresponding to 0.5 to 8.0 keV.
Spectra were binned as needed to obtain at least 20 counts per fitting bin
 (before background subtraction) to ensure applicability of the $\chi^2$
 statistic.
For sources resulting in fewer than 5 degrees of freedom, the
 unbinned spectra (without background subtracted) were also fit using the
 C-statistic to check for fit parameter consistency.
Redistribution matrices and ancillary response files were generated using
 \cxo\ X-ray Center tools and calibration data (CIAO version 2.3) using the
 most recent gain maps and
 observation-specific bad pixel lists and aspect histories.

For sources imaged on front-side illuminated ACIS CCDs, the
 algorithm developed by Townsley \etal\ (2000) was applied to partially
 correct for the effects of charge loss and smearing caused by
 charge-transfer inefficiency.
The algorithm was applied to the Level~1 event list, a new Level~2 file
 generated (following the CIAO threads), and the spectra re-extracted.

The {\tt acisabs}\footnote{http://www.astro.psu.edu/users/chartas/xcontdir/xcont.html} correction was applied
 (as a multiplicative model
 in XSPEC) in all spectral fits to account for the
 temporal decrease in the low-energy sensitivity of the ACIS detectors.
Sources with detected count rates in excess of 0.1 c~frame$^{-1}$
 were treated for effects of pileup following the
 procedures described in Davis (2002) and using the event pileup
 model of Davis (2001).
No CTI-correction was applied to piled-up spectra.

For each model, the fit parameters, 0.5--8.0~keV model flux, $\chi^2$
 statistic and number of degrees of freedom, and corresponding errors were
 recorded.
All errors are extremes on the single interesting parameter 90\%
 confidence intervals except for derived fluxes (and corresponding luminosities)
 where the 1$\sigma$ errors are quoted.
XSPEC-determined fluxes (see \S~\ref{s:pegpwrlw}),
 averaged over the duration of the observation,
 were then scaled by the inverse of the fraction of
 the telescope point spread function (PSF)
 within the source extraction region and luminosities
 computed using the adopted distances to the host galaxies.

X-ray sources in each galaxy field were sorted by number of detected counts.
Spectral fits were then made beginning with the highest-count sources
 and continuing with lower-count sources until all potential ULX candidates
 were examined.
This resulted in spectral fits for many sources with luminosities
 $<$$10^{39}$~\ergl.

In addition to fitting models to the 0.5-8.0 keV spectrum of the highest-count
 sources, the background-subtracted X-ray counts for all extracted
 sources were binned into three broad bands;
 defined as $S$ (0.5-1.0 keV), $M$ (1.0-2.0 keV), and $H$ (2.0-8.0 keV);
 and the X-ray colors $(M-S)/T$ and $(H-M)/T$, where
 $T=S+M+H$, were constructed following Prestwich \etal\ (2003).

\subsubsection{Determining ULX Status Spectroscopically} \label{s:pegpwrlw}

By our definition, a ULX is a discrete source whose
 {\em intrinsic} luminosity exceeds
 $10^{39}$~\ergl\ in the 0.5--8.0 keV energy band.
An absorbed (multiplicative
 {\tt acisabs} and {\tt phabs} components) XSPEC-specific power law model called
 {\tt pegpwrlw} was applied
 to the X-ray spectra to obtain the intrinsic luminosity.
The normalization parameter
 for this  model is the intrinsic flux in a prescribed bandpass.
The uncertainty estimate for this parameter then provides the uncertainty in
 the intrinsic flux.
The {\em observed} flux and corresponding uncertainty were
 obtained using XSPEC's {\tt flux} command.

The power law model provides a statistically-acceptable fit to most
 of the spectra examined (\S \ref{s:RESULTS_timing}).
For these spectra, the
 average value of the quantity $(\Delta F/F)\sqrt{N} \sim 1.7\pm0.3$;
 where $F$ is the intrinsic flux, $\Delta F$ its
 uncertainty, and $N$ the number of source counts.
For spectra poorly fit using an absorbed power law, the intrinsic flux was
 estimated using the {\tt flux} command
 applied to the best-fitting model with the model absorption component
 temporarily set to zero and the uncertainty was estimated as
 $\Delta F = 1.7(F/\sqrt{N})$.

 \subsubsection{Temporal Analysis}

From the time series of events within the source extraction regions,
 the X-ray light curves were grouped into 1000-s bins and
 $\chi^2$ tests performed against the constant count rate hypothesis to test for
 source variability.
The Kolmogorov-Smirnov (K-S) statistic was also computed to test the sources
 for time variabiliy by comparing the cumulative event arrival times, binned
 at the nominal frame time (3.24~s in full-frame mode), to that expected
 for a steady source.

For the subset of variable sources, as deduced from either of these two tests,
 power density spectra (PDS) were also examined.
Using the Leahy normalization (Leahy \etal\ 1983),
 the average power is 2.0 (as expected from Poisson noise) and
 fluctuations up to 10--15 are commonly seen
 with a maximum power in one or two frequency
 bins typically between 15 and 20 in the normalized power spectra.
Fluctuations of this order are due to noise.
No values above these levels were observed.
Therefore, no statistically-significant periodicities were obtained from
 analysis of the power spectra.

\subsubsection{Spatial Analysis}

The data were checked to ensure the nominal pointing
accuracy\footnote{http://asc.harvard.edu/cal/ASPECT/fix\_offset/fix\_offset.cgi}
 using the latest alignment files (c. 2002 May 02).
The accuracy of absolute positions in the \cxo\ data
 have a typical rms radius of 0.\arcsec 6
 (Aldcroft \etal\ 2000).
Refinements to absolute positions conceivably can be made
 by cross-correlating X-ray
 positions with available astrometric catalogues.
However, this has proved impractical for the following reasons:
(1) By design, analysis is restricted to
 the small, source-crowded, areas of the sky within the host
 galaxy's \elip.
(2) The sample of galaxies are typically far from the
 Galactic plane where field stars are relatively rare.

Refined estimates of the centroids of the sources were made by fitting
 an elliptical Gaussian to the spatial distribution of X-ray
 events.
This method works well for locating the
 centroids of the model \cha\ point spread functions, at various off-axis angles
 and (monochromatic) energies, available from the
 \cxo\ X-ray Center PSF 
library\footnote
{ftp://cda.harvard.edu/pub/arcftp/caldb/acis\_psflib\_2.9.tar}.
For ULX candidates (those with of order 100~c or higher), we estimate the
 statistical uncertainty in the source positions due to centroiding errors
 is less than 0.\arcsec 1.
Source celestial positions quoted in this work are these centroid-refined
 positions.

Radial profiles were extracted from a subset of sources and compared to model
 PSFs available from the \cxo\ X-ray Center.
The subset of sources was chosen as those with $>$10\% of X-ray events
 within the source extraction region attributable to background.
As the background region encircles and is contiguous
 with the source region, sources with high background are either embedded in
 diffuse emission (a true background) or are extended.
Where other sources were nearby, these were deleted from the data prior to
 extracting the radial profile.
No ULX candidate appeared significantly extended according to this comparison.


\section{RESULTS} \label{s:RESULTS}

The resulting sample of galaxies and some of their properties
 (distance, \lb, \lfir, and nearest-neighbor distance $\theta_p$)
 are listed in Table~1.
Also included is a log of the corresponding \cxo\ observations.
In a few cases, two \cxo\ observations of a single galaxy were combined to
 increase the signal-to-noise but
 each observation is listed separately in Table~1.
The observation log includes the fraction, $f_{\rm FOV}$, of the angular size
 defined by the \elip\ ellipse falling within the instrument field of view;
 an estimate, $L_{min}$, of the luminosity of the weakest source
 for which a statistically-meaningful spectral fit could be made;
 and citations to previous analysis of the X-ray data from the literature.
Observations performed in sub-array mode are noted in the $f_{\rm FOV}$
 column of Table~1.
The tabulated luminosity, $L_{min}$, corresponds to roughly 50 source counts.
The source {\sl detection} limit is about 10~c.

\subsection{The Sample of Galaxies} \label{s:RESULTS_sample}

The distribution of galaxies in the \cxo\ sample over Hubble morphological
 type is shown in Figure~\ref{f:hubble_type}.
There are 18 elliptical galaxies, 9 lenticulars, 46 spirals, and 9
 irregulars and peculiars for a total of 82 galaxies.
Typically, in this work, galaxies earlier than S0/a are
 collectively referred to as ellipticals and the remainder
 as spirals.
While the full range of morphological types is represented in the \cxo\
 sample, the 1:2 ratio of ellipticals-to-spirals is higher than the
 ratio 1:4 in the Tully (1988) Nearby Galaxy Catalogue
 and the $\sim$1:3 ratio in the RC3.

\begin{center}
\includegraphics[angle=-90,width=\columnwidth]{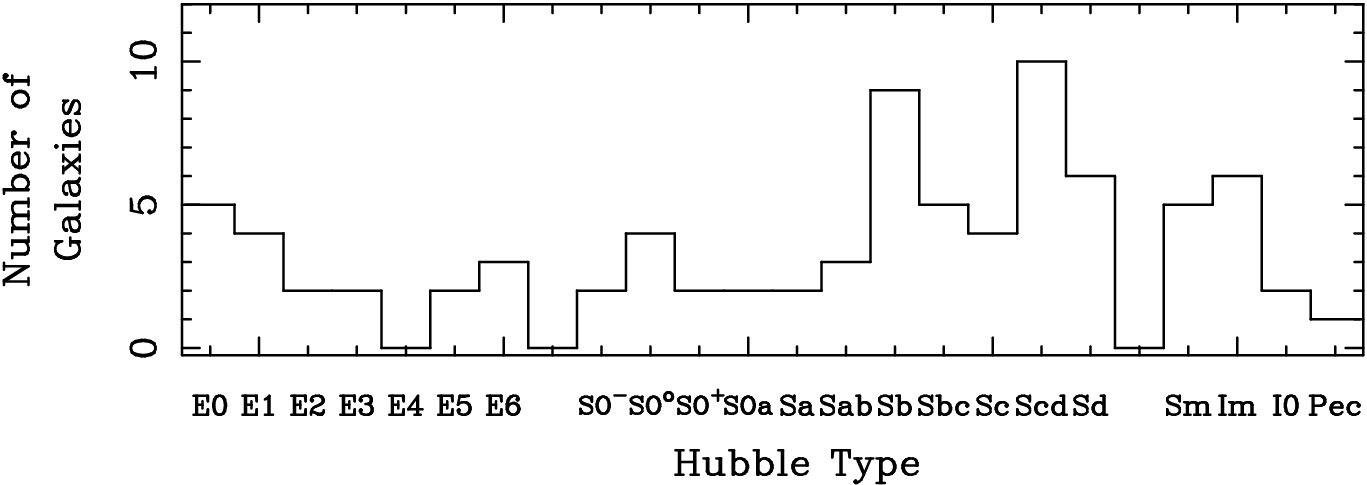}
\figcaption{Distribution of Hubble morphological types among
 the \cxo\ sample of galaxies. \label{f:hubble_type} }
\end{center}

The \cxo\ sample of galaxies is not expected to be complete to any
 specified limiting distance, magnitude, or other property.
However, the \cxo\ sample can be compared to, for example, the Tully (1988)
 Nearby Galaxy Catalogue of 2367 galaxies.
The distribution of absolute blue magnitudes of the \cxo\ sample of
 galaxies is compared in Figure~\ref{f:cxo_tully.B}
 to those of the 1768 galaxies in the Tully catalogue with tabulated
 absolute blue magnitudes.
The average of the \cxo\ sample is 0.8 mag brighter in blue light than is the
 Tully sample.
The distributions of galaxy distances for the \cxo\ sample and for the Tully
 catalogue are shown in Figure~\ref{f:cxo_tully.D}.
The average distance of galaxies in the \cxo\ sample is 11.5$\pm$1.2~Mpc
 compared to 22.5$\pm$0.4~Mpc for the Tully galaxies.
Similarly, the \elip\ angular sizes are larger for the \cxo\ sample
 (average $8.\arcmin 2 \pm 0.\arcmin 8$)
 compared to the Tully galaxies
 ($3.\arcmin 8 \pm 0.\arcmin 5$).
Thus, the \cxo\ sample is comprised of nearby galaxies of large
 angular size and high blue luminosity compared to the Tully catalogue of
 galaxies.
The sample is also expected be more luminous in X-ray light as
 X-ray luminosity correlates with blue luminosity
 (e.g., Fabbiano \& Trinchieri 1985; Fabbiano 1989) though the correlation
 varies with morphological type and is nonlinear (Fabbiano \& Shapley 2002).
The departure of the \cxo\ sample from the Tully catalogue
 is not unexpected as the \cxo\ sample represents an X-ray-selected sample.

\begin{center}
\includegraphics[angle=-90,width=\columnwidth]{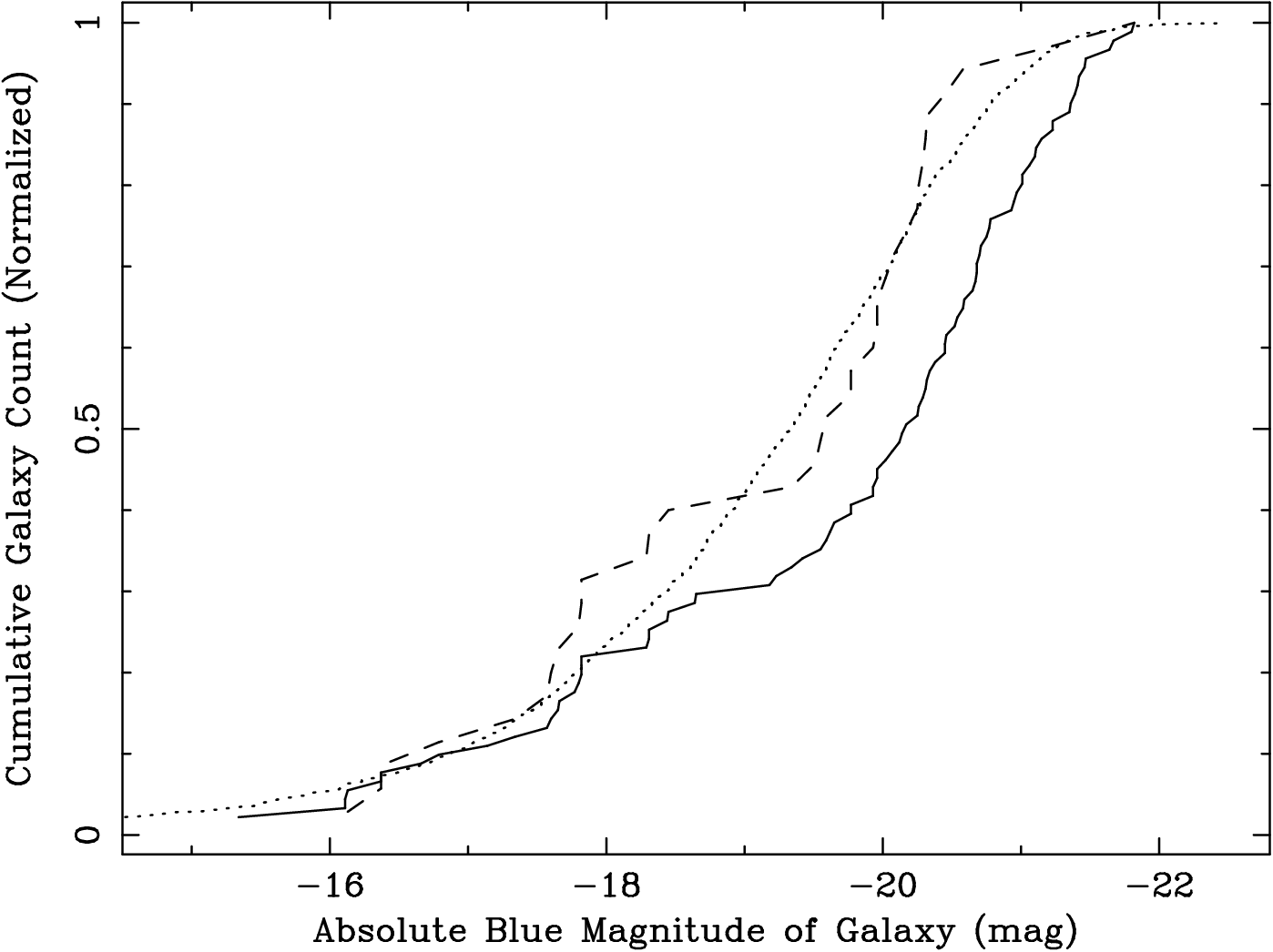}
\figcaption{Cumulative distribution of the \cxo\ sample of galaxies
({\sl solid} curve) compared to galaxies from the Tully (1988) catalogue
({\sl dotted}) over absolute blue luminosity. Also shown ({\sl dashed}) is
the cumulative distribution of a subsample of the \cxo\ galaxies having the same
distribution of absolute blue luminosity as the Tully catalogue.
\label{f:cxo_tully.B} }
\end{center}

\begin{center}
\includegraphics[angle=-90,width=\columnwidth]{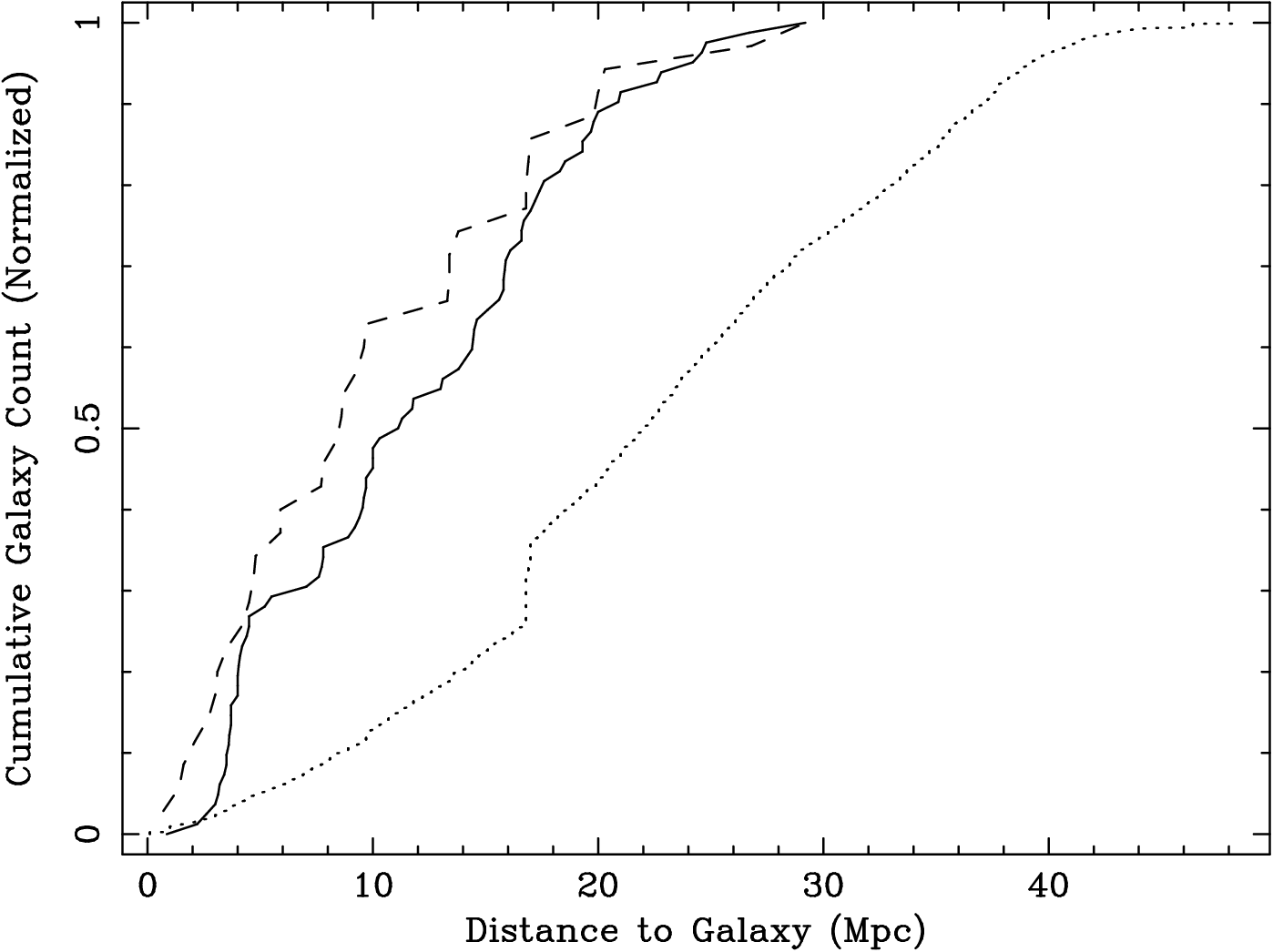}
\figcaption{Cumulative distribution of the \cxo\ sample ({\sl solid} curve) and
subsample ({\sl dashed}, see text) of galaxies compared to galaxies
from the Tully (1988) catalogue ({\sl dotted}) over distance.
\label{f:cxo_tully.D} }
\end{center}

In order to further assess the extent of any selection bias in the \cxo\ sample,
 a subsample of galaxies having the same distribution of absolute $B$
 magnitudes as Tully-catalogued galaxies within $D$$=$29~Mpc was selected.
In \S~\ref{s:hosts}, this subset is used
 along with the full \cxo\ sample of galaxies to study correlations between the
 ULX candidates and properties of their host galaxies.
The subsample was chosen by first grouping the full sample and the Tully
 catalogue into 4 bins spanning -22.5$<$M$_{\rm B}$$<$-14.5 with a 2.0~mag
 width for each bin.
(There were too few galaxies in the full \cxo\ sample to afford a smaller
 binsize.)
The subsample was then constructed by randomly selecting galaxies from each
 bin such that the fraction of subsample galaxies in each bin equals the
 fraction of Tully galaxies in that bin.
The distribution of absolute blue magnitudes and distances of the subsample
 are shown in Figures~\ref{f:cxo_tully.B} and~\ref{f:cxo_tully.D}, respectively.
The average subsample galaxy distance is 10.2~Mpc and the average \elip\
 angular size is 9.\arcmin 3.
The Tully catalogue of galaxies within 29~Mpc and with tabulated $M_{\rm B}$
 average 16.4~Mpc and 4.\arcmin 2.

The $(B-V)$ colors of the \cxo\ sample all fall within the range of 0.5 to 1.0
 mag.
Therefore, although blue light is more sensitive to the properties of
 the stellar population of the host galaxy than is visible light, the blue
 luminosity is used here as a proxy for galaxy mass with an estimated factor of
 two uncertainty assuming ${\rm M}_{\rm V} \propto$~mass.
On the other hand, a galaxy's far-infrared luminosity is proportional to its recent star formation rate (Kennicutt 1998).
The distribution of the sample of galaxies in blue and far-infrared luminosity
 is shown in Figure~\ref{f:BvF}.
Elliptical galaxies tend to cluster near the upper left in this figure
 indicative of a relatively high mass but little or no
 recent star formation.
In contrast, spiral galaxies  span from lower-left to upper-right
 showing a general trend of star formation rate proportional to galaxy mass.
It is also evident that spirals have a higher star formation rate per unit
 mass (blue light) compared to the ellipticals.

\begin{center}
\includegraphics[angle=-90,width=\columnwidth]{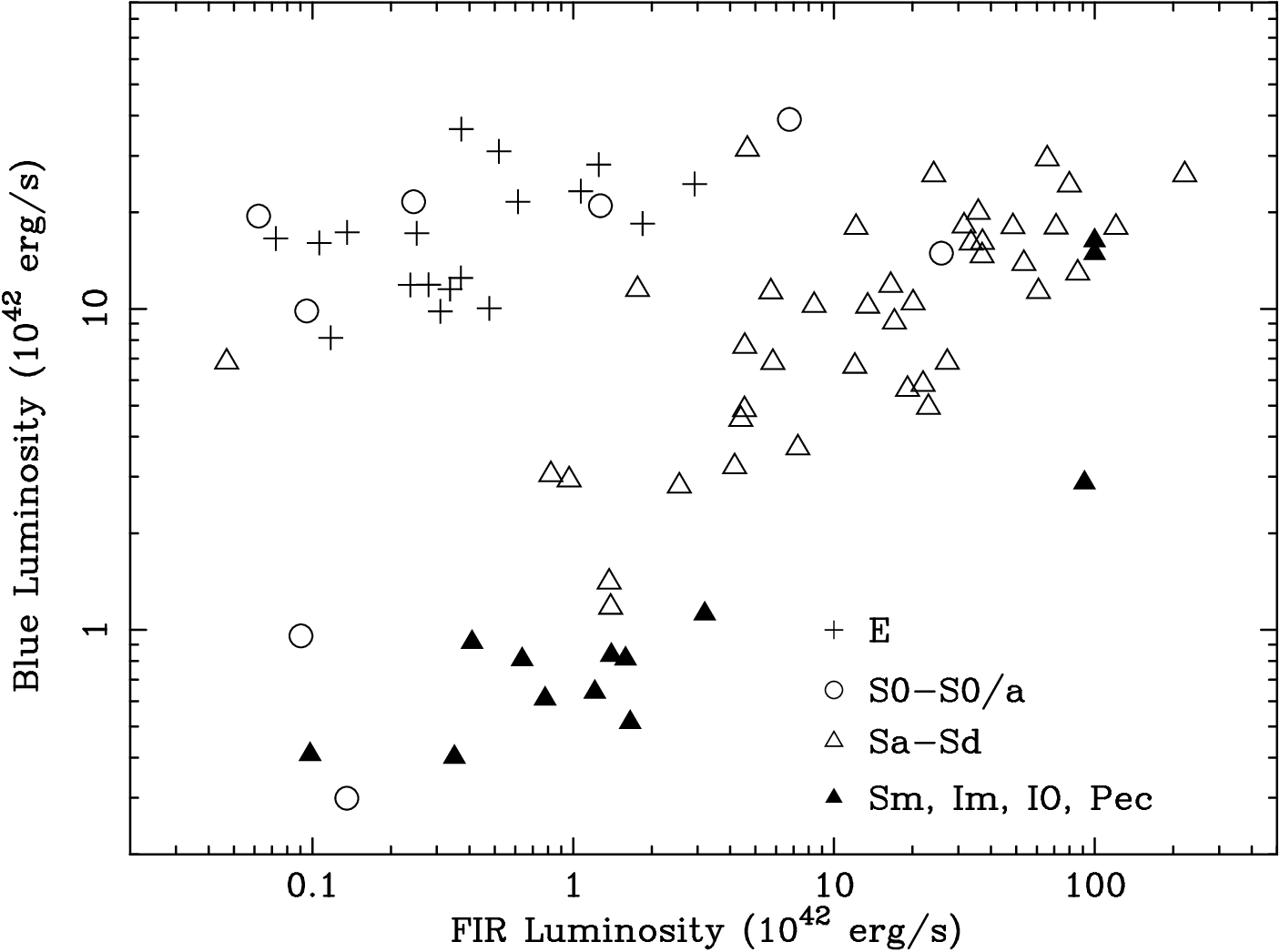}
\figcaption{Distribution of the \cxo\
sample of galaxies in blue and far-infrared luminosity. \label{f:BvF} }
\end{center}

\subsection{The ULX Candidates} \label{s:RESULTS_ulx}

Approximately 3400 X-ray sources were detected above a signal-to-noise
 ratio of 2.8 (resulting in $>$10~c per source and much less than 1 false
 detection per field) within the \elip\ areas of the 82 galaxies analyzed.
Among these are $\sim$1900 sources with intrinsic luminosities
 estimated from detected counts exceeding $\sim$$10^{38}$~\ergl.
 The source-region image, spectrum, and light curve were examined for
 all these sources though only 837 had sufficient counts ($>$50)
 for further analysis.
Detailed spectral fits were made to 357 of these sources and 154 were
 determined to have luminosities in excess of $10^{39}$~\ergl\ in the
 0.5--8.0~keV energy range.

We may therefore define three source populations: (1) The 837-member group
 above an estimated $10^{38}$~\ergl\ and with $>$50~c detected for which
 X-ray colors have been determined and the K-S test applied to the
 X-ray light curves,
 (2) The 357-member subset for which spectral parameters and hence luminosities
 have been established, and (3) the subset of 154 ULX candidates.
Table~2 lists the celestial positions and X-ray properties of the ULX
 candidates.

The subset of ULX candidates is complete to the extent that the derived
 luminosities (and hence adopted distances) are correct.
The completeness level of the 357 sources with spectral fits made
 can be estimated as
 the luminosity where their cumulative luminosity function begins to flatten.
This occurs at \lx $\sim$$4\times 10^{38}$~\ergl\ with about 20\%
 lying below this completeness level.
These lower-luminosity sources tend to be those from nearby galaxies with
 deep exposures but are otherwise not expected to differ from sources above
 the completeness limit.
The lowest luminosity of a source for which spectral parameters have been
 established is $8.9 \times 10^{37}$~\ergl.
The luminosities for the first group were estimated from the number of
 detected counts in the manner described in \S~\ref{s:METHODS_galSelect} and
 are therefore much more uncertain than for the other two source populations.
The completeness level for this population, based on its
 cumulative luminosity function is also $\sim$$4\times 10^{38}$~\ergl.
More importantly for this population, however, is that there are sufficient
 counts detected for analyzing their X-ray colors and timing properties
 without overwhelming statistical uncertainties. Thus, this population is
 limited to sources with $>$50~c detected.

\subsubsection{X-ray Spectra} \label{s:RESULTS_spectra}

The 3-parameter absorbed power law model provides one of the simplest
 phenomonological descriptions of X-ray spectra.
This model was applied to the
 observed spectra to constrain the observed and intrinsic luminosities
 of the sources (see also \S~\ref{s:pegpwrlw} above)
 and to establish their basic spectral shapes.
Statistically-acceptable ($>$90\% confidence) fits were obtained using this
 model for 83\% (298 of 357) of sources for which spectral parameters
 have been determined and for 130 (84\%) of the ULX candidates.
This does not imply that a power law is the only acceptable model for these
 spectra
 nor that a power law fully describes the physical mechanism(s) responsible for
 the observed X-ray emission.
Indeed, an absorbed thermal emission line ({\tt mekal}) model, for example,
 results in an acceptable fit to $\sim$78\% of the sources analyzed.

The fitted power law indices of the ULX candidates with statistically-acceptable
 power-law fits are compared in
 Figure~\ref{f:binGamma} to the weaker sources.
Overall, the distributions are very similar and can be described by the
 same Gaussian shape.
However, there is a significant population of sources (20 of 130)
 with $\Gamma$\GA 3 in the ULX population that does not have a
 counterpart in the sample of weaker sources (7 of 168).

Closer inspection showed that
sources with a steep power law index often also required a high absorbing
 column density that compensates for an observed spectral rollover
 toward low energies.
The average column density is $N_H/10^{20}$$=$$38.7\pm3.0$~cm$^{-2}$
 for ULX candidates with $\Gamma$$>$3 compared to $19.6\pm1.2$ for the
 $\Gamma$$\le$3 sources.
The ultimate effect is to promote steep power law
 sources with low {\sl observed} luminosities
 into the high {\sl intrinsic} luminosity ULX candidate category.
In contrast, thermal models such as the {\tt mekal} model have
 a natural rollover at low energies, do not require as large an
 absorbing column density, and result in a lower intrinsic luminosity.

\begin{center}
\includegraphics[angle=-90,width=\columnwidth]{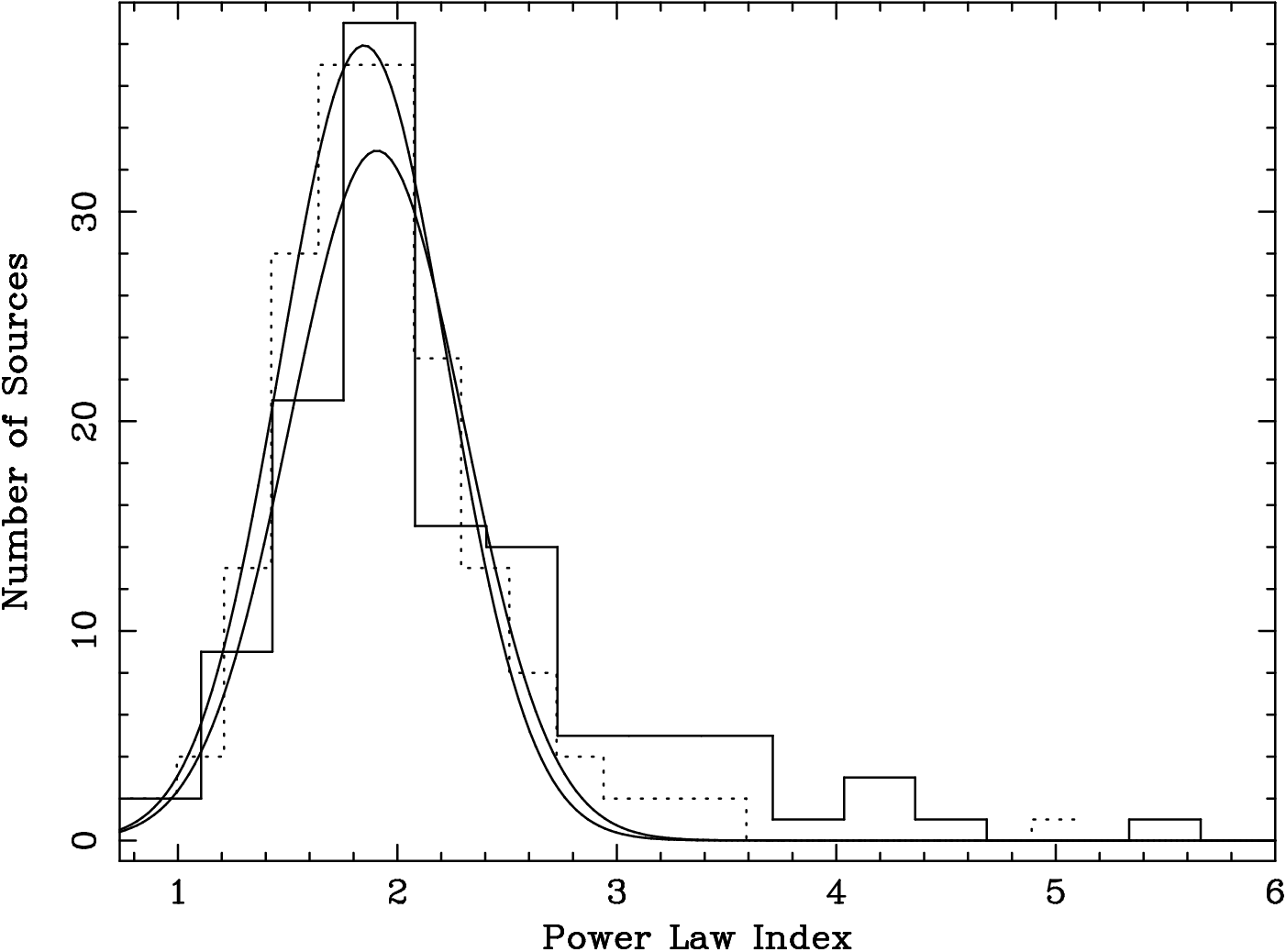}
\figcaption{Number of ULX candidates ({\sl solid} histogram) and non-ULX
sources for which spectral fits were made ({\sl dotted}) against
power law index. The two distributions include only sources for which an
absorbed power law model provides an acceptable fit. Both distributions
can be described by a Gaussian with
the same fit parameters within errors
(centroids 1.97$\pm$0.11 and 1.88$\pm$0.06 for the ULXs and the other
sources, respectively, and widths 0.50$\pm$0.10 and 0.41$\pm$0.05).
The population of ULX
candidates with power-law index $\Gamma$$>$3 is discussed in the text.  \label{f:binGamma} }
\end{center}

This trend is illustrated in the upper panel of
 Figure~\ref{f:LmLpvsGamma} where the ratio of
 the intrinsic luminosity derived using the {\tt mekal} model to
 that derived using the power law model is shown against the power law
 index. Only a subset of ULX candidates are included in
 this figure: A sample with $>$500 detected counts
 and $\Gamma$$<$3 and all of the $\Gamma$$>$3 sources though,
 in both domains, only sources with statistically-acceptable fits from both the
 power law and the {\tt mekal} model are included.
The difference in derived intrinsic luminosities increases rapidly with
 increasing $\Gamma$.
For the bulk of the sources modeled
 (those with $\Gamma$$\sim$1.8) however, both the {\tt mekal} and power law
 models result in very similar intrinsic luminosities.

\begin{center}
\includegraphics[angle=-90,width=\columnwidth]{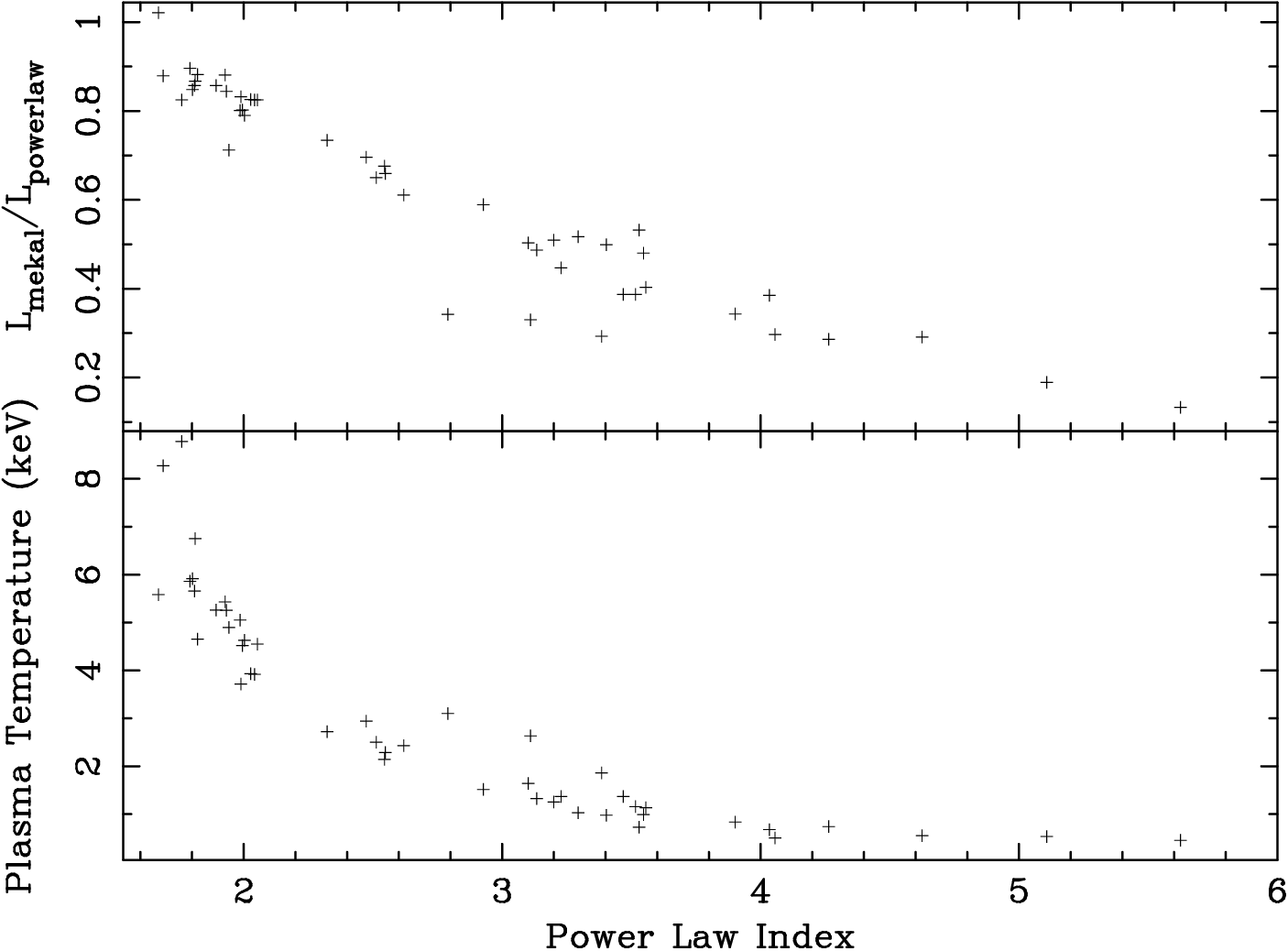}
\figcaption{({\sl Upper panel}) Ratio of
 the intrinsic luminosity derived using the {\tt mekal} model to
 that derived using the power law model is shown against the power law
 index. The power law model predicts a higher intrinsic luminosity
 compared to the {\tt mekal} model with the difference systematically
 increasing with power law index. ({\sl Lower panel}) Corresponding
 {\tt mekal} model plasma temperature decreases with increasing $\Gamma$.
 \label{f:LmLpvsGamma} }
\end{center}

This systematic effect helps explain the high-$\Gamma$ tail apparent in
 the ULX candidate population of Figure~\ref{f:binGamma} but lacking in the
 lower luminosity source population.
Applying the {\tt mekal} model to the 20 ULX and
 7 non-ULX sources with $\Gamma$$>3$ and acceptable power law fits resulted in
 {\tt mekal}-derived intrinsic luminosities $<$$10^{39}$~\ergl\ for 17 of
 the 27 sources. These 17 are denoted in Table~2.
Thus, adopting the {\tt mekal} luminosities would result in
 similar high-$\Gamma$ distributions for both populations.
The lack, at the outset, of steep power law sources in the non-ULX
 population is due to the relatively low number of counts in the 0.5--8.0 keV
 band for this spectral shape combined with the fact that
 low-count sources were not selected for spectral analysis.
However, the
 X-ray color analysis of \S~\ref{s:RESULTS_colors} shows that there are
 sources with colors equivalent to a steep power law in the
 low-count population.

It is evident that more than one spectral model often provides
 statistically-acceptable fits to the observed X-ray spectra.
Often these models ostensibly represent quite different physical
 emission mechanisms.
For example, physical models of X-ray phenomena that predict
 ``soft'' X-radiation, corresponding to a steep power law index,
 are cool thermal emission sources (lower panel, Figure~\ref{f:LmLpvsGamma}),
 such as supernovae,
 while phenomena that can be characterized as non-thermal, such as synchrotron
 emission from a power-law distribution of electrons or Comptonization,
 tend to result in relatively flat power law indices.
Thus, while either model is statistically-acceptable,
 it is reasonable to {\sl assume} that the majority of the steep power law
 sources are thermal sources.
Many of these may be supernovae though
 three $\Gamma$$>$3 sources are among the most variable X-ray
 sources in the ULX sample (Table 2 and \S~\ref{s:RESULTS_timing}).
One of these is a known super-soft source in NGC~5457 (Mukai \etal\ 2003).

Having established that both the power law and the {\tt mekal} models
 predict similar luminosities for the majority of sources (those near
 $\Gamma$$\sim$1.8) and that both models effectively disriminate between
 ``soft'' and ``hard'' X-ray spectra, it remains to chose one or both
 models (or some other) as a baseline.
The power law is arguably the easier to grasp conceptually and requires
 the fewest number of parameters.
Therefore, the absorbed power law model fit parameters and
 derived luminosities are reported in Table~2 whenever this model
 provided a statistically-acceptable fit.
Additional models applied to sources
 not well-fit by an absorbed power law were
 thermal emission line models ({\tt mekal} or {\tt vmekal}),
 disk blackbody models ({\tt diskbb}) and combinations of these three (including
 power law) basic spectral models.
Of the 24 ULX candidates poorly fit by the absorbed power-law model, 11 were
 acceptably-fit using 1-- or 2--component
 thermal emission line models, 9 using disk blackbody models,
 and 4 were not well-fit by any of the trial spectral models.
Parameters for these models are reported in Table~2 (the power law parameters
 are quoted for the 4 spectra with no acceptable fits).

Figure~\ref{f:GammavsCts} shows the value of the power law spectral index,
 $\Gamma$,
 against the number of detected source counts and Figure~\ref{f:GammavsL}
 displays $\Gamma$ against the derived intrinsic source luminosity.
Not surprisingly,
 sources for which the simple absorbed power law model was unacceptable
 tend to be those sources with a high number of detected counts
 (e.g., a power law is unacceptable for 39\% of the 51 sources with
 $>$1000 detected counts compared to only 8\% of 177 sources with $<$200
 detected counts).

\begin{center}
\includegraphics[angle=-90,width=\columnwidth]{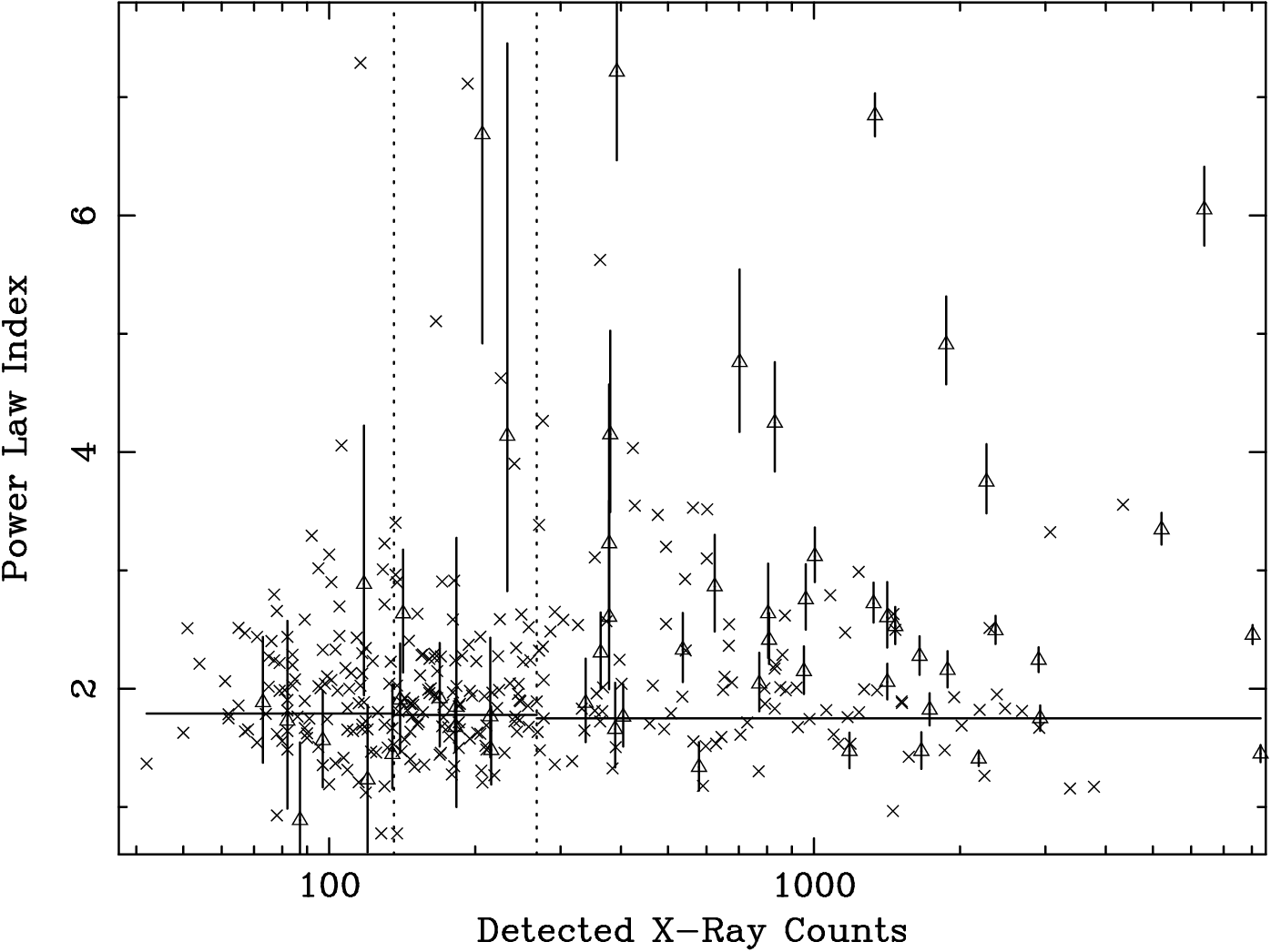}
\figcaption{Results of simple absorbed power law model fits.
Sources for which this model provides a
statistically-acceptable fit at the 90\% confidence level are marked with
an $\times$ and those unacceptably-fit by a simple power law are denoted
by triangles.
Errors shown are extremes on the single interesting parameter 90\%
 confidence intervals.
They indicate typical values for all sources with power law models applied but
 are not displayed for sources with {\sl acceptable} fits for clarity.
Formal errors for the ULX candidates are provided in Table~2.
Horizontal lines denote the average power law index, for sources with an
 acceptable fit, on domains spanning an equal number of sources. For
 clarity, these domains are separated by dotted vertical lines.
 \label{f:GammavsCts} }
\end{center}

\begin{center}
\includegraphics[angle=-90,width=\columnwidth]{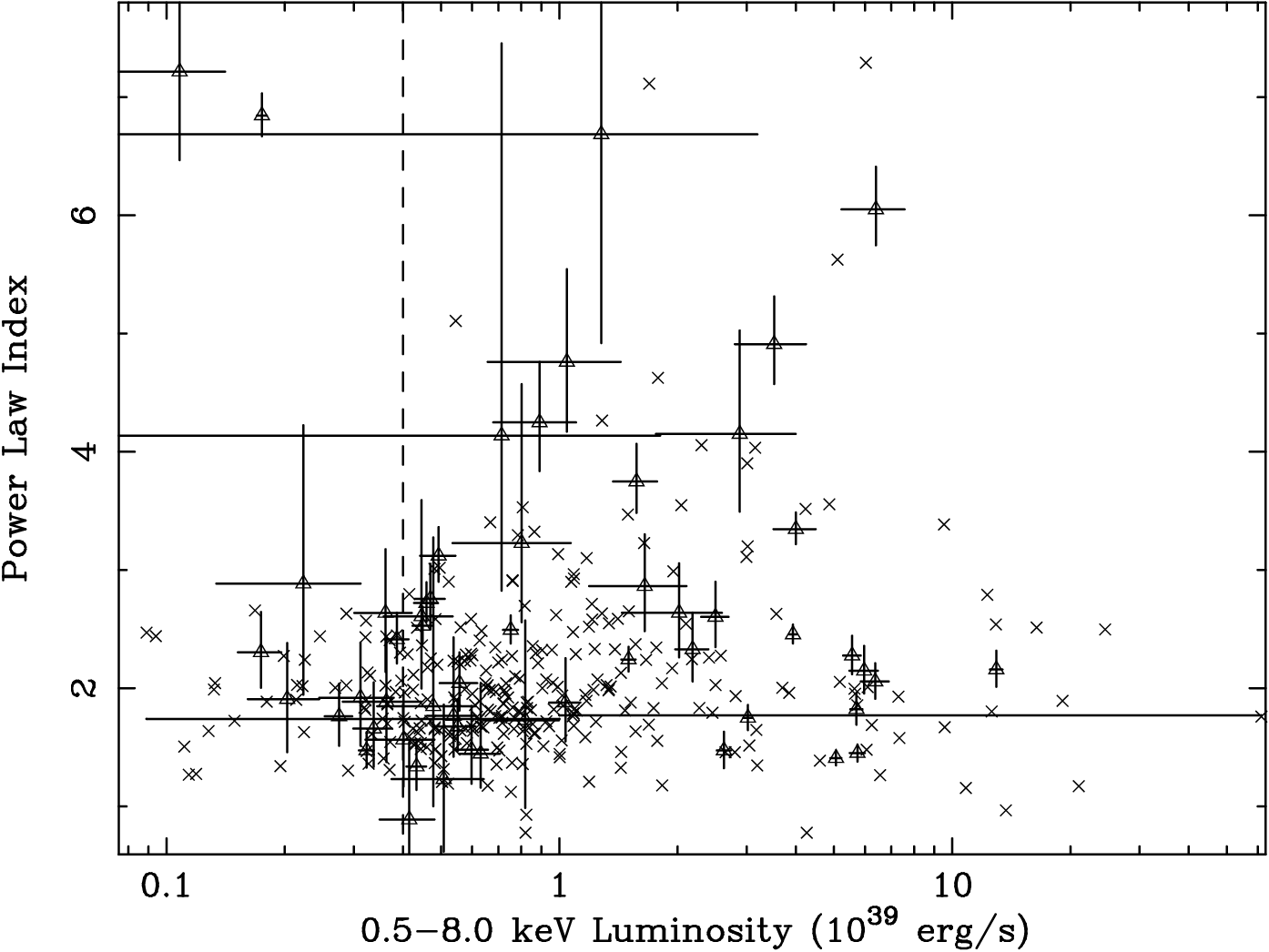}
\figcaption{Results of simple absorbed power law model fits.
Symbols are the same as in Figure~\ref{f:GammavsCts}.
Note sources poorly-fit with a power law are uncorrelated with
intrinsic luminosity and that many of the highest luminosity sources are
well-fit with this simple model.
Horizontal lines denote the average power law index, for sources with an
 acceptable fit, for ULX and for weaker sources.
The vertical line denotes the estimated completeness limit of the sample of
 sources for which spectral modeling has been performed.
\label{f:GammavsL} }
\end{center}

The average value of the power law index,
 $\langle \Gamma \rangle$, for sources with
 statistically-acceptable power law fits,
 are 1.79$\pm$0.09 (for sources with $<$137~c), 1.78$\pm$0.07
 (137 to 268~c), and 1.75$\pm$0.03 ($>$268~c) where the 3 ranges were
 chosen to contain equal numbers of sources.
Similarly, $\langle \Gamma \rangle$ is displayed in Figure~\ref{f:GammavsL}
 for the ULX candidates (1.74$\pm$0.03), and for the weaker sources
 (1.77$\pm$0.04 above the completeness limit and 1.76$\pm$0.10 for sources
 below $4\times 10^{38}$~\ergl).
The dispersion of $\Gamma$ about the mean value is $\sigma$$=$1.18 for the
 ULX candidates and 0.43 for the non-ULX sources.
The steep power law sources account for the larger dispersion in the
 ULX candidates population.
Restricting the range to
 $\Gamma$$<$3 results in $\sigma$$=$0.51 and 0.40 for the two groups.
The average power law index is therefore independent of the
 number of detected counts and of the intrinsic luminosities
 of the source above our completeness limit of $\sim$$4\times 10^{38}$~\ergl.
\cxo\ studies of the nearby galaxies, e.g., M31 (Kong \etal\ 2002) and M81
 (Swartz \etal\ 2003), conclude that this trend continues
 ($\langle \Gamma \rangle \sim 1.8$) even for
 sources as weak as $3.4\times 10^{37}$~\ergl\ and $2\times 10^{37}$~\ergl,
 respectively.

\subsubsection{X-ray Colors} \label{s:RESULTS_colors}

There are many more weak sources detected than can be meaningfully
 analyzed through spectral fitting. A common tool used for low-count
 sources and low-resolution spectra is the X-ray color-color diagram.
Prestwich \etal\ (2003) advocate $(H-M)/(H+M+S)$ and
 $(M-S)/(H+M+S)$ as colors that provide some physical insight into the
 nature of X-ray source populations and these colors are adopted here.

The X-ray colors of the 154 ULX candidates and of the 837 sources with estimated
 luminosities above $10^{38}$~\ergl\ and $>$50~c
 are shown in Figures~\ref{f:colorU} and~\ref{f:colorAll}, respectively.
Error bars, propagated from the statistical uncertainties in the three
 X-ray bands, are omitted from Figure~\ref{f:colorAll} for clarity.
Colors of absorbed power law model spectra (for sources imaged on
 back-illuminated CCDs) are also shown for reference as solid curves.
The front-illuminated devices have relatively lower response in the
 soft band so that similar curves for these
 devices do not extend to such low values of $(M-S)/T$.
Nevertheless, the differences are typically much less than the uncertainties
 in the X-ray colors.

\begin{center}
\includegraphics[angle=-90,width=\columnwidth]{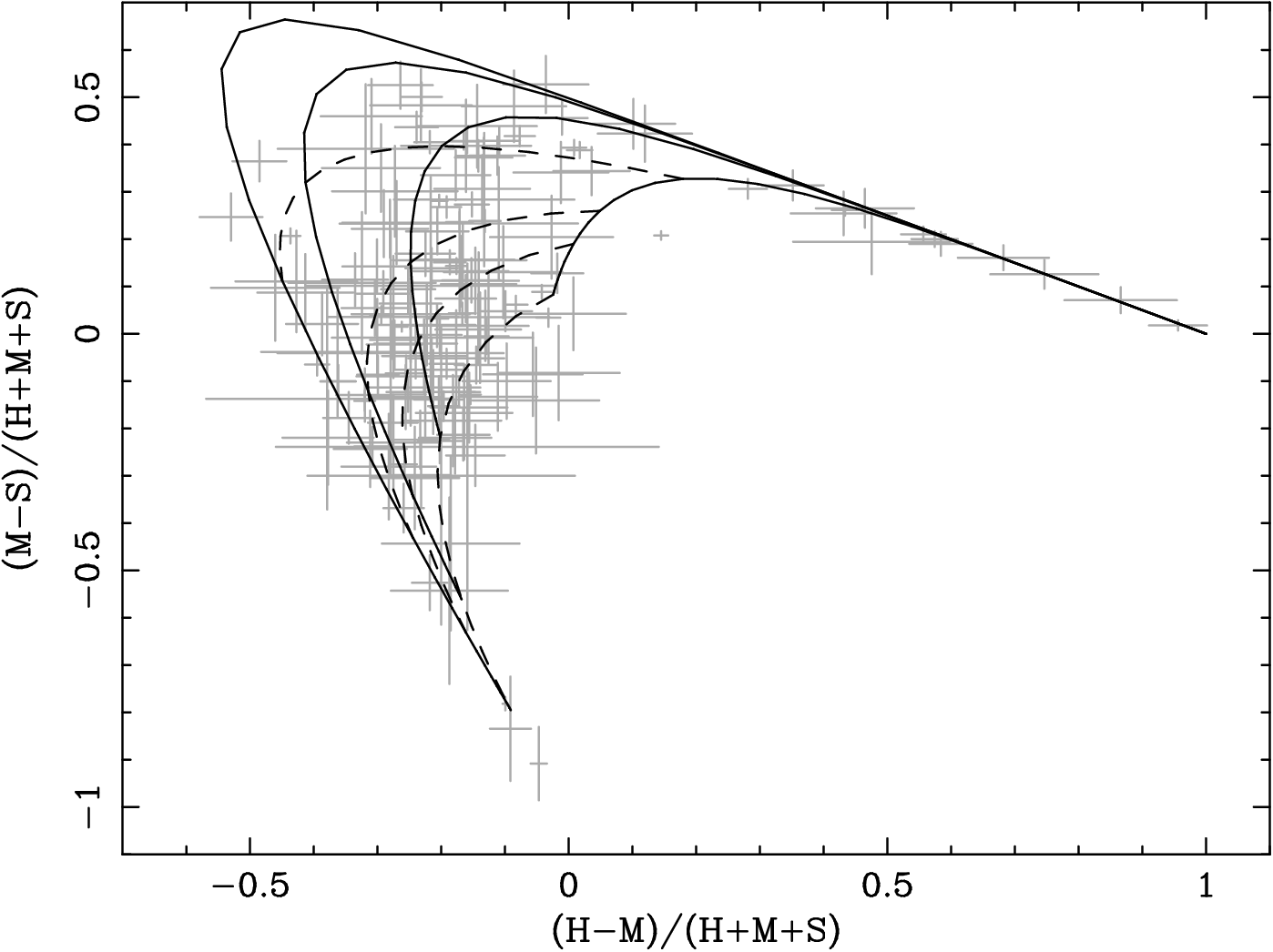}
\figcaption{X-ray colors of the ULX candidates.
Solid curves denote colors of absorbed power law models of spectral indices
$\Gamma=1$, 2, 3, and 4  (from right to left) and for the range of absorbing
columns $n_H=10^{20}$ to $10^{24}$~cm$^{-2}$. Dashed curves denote constant
absorption columns of $n_H = 10^{20}$, $10^{21}$, $2\times 10^{21}$, and
$5\times 10^{21}$~cm$^{-2}$ (from bottom to top).
Errors shown were propagated from the statistical uncertainties in the three
 X-ray bands.
 \label{f:colorU} }
\end{center}

\begin{center}
\includegraphics[angle=-90,width=\columnwidth]{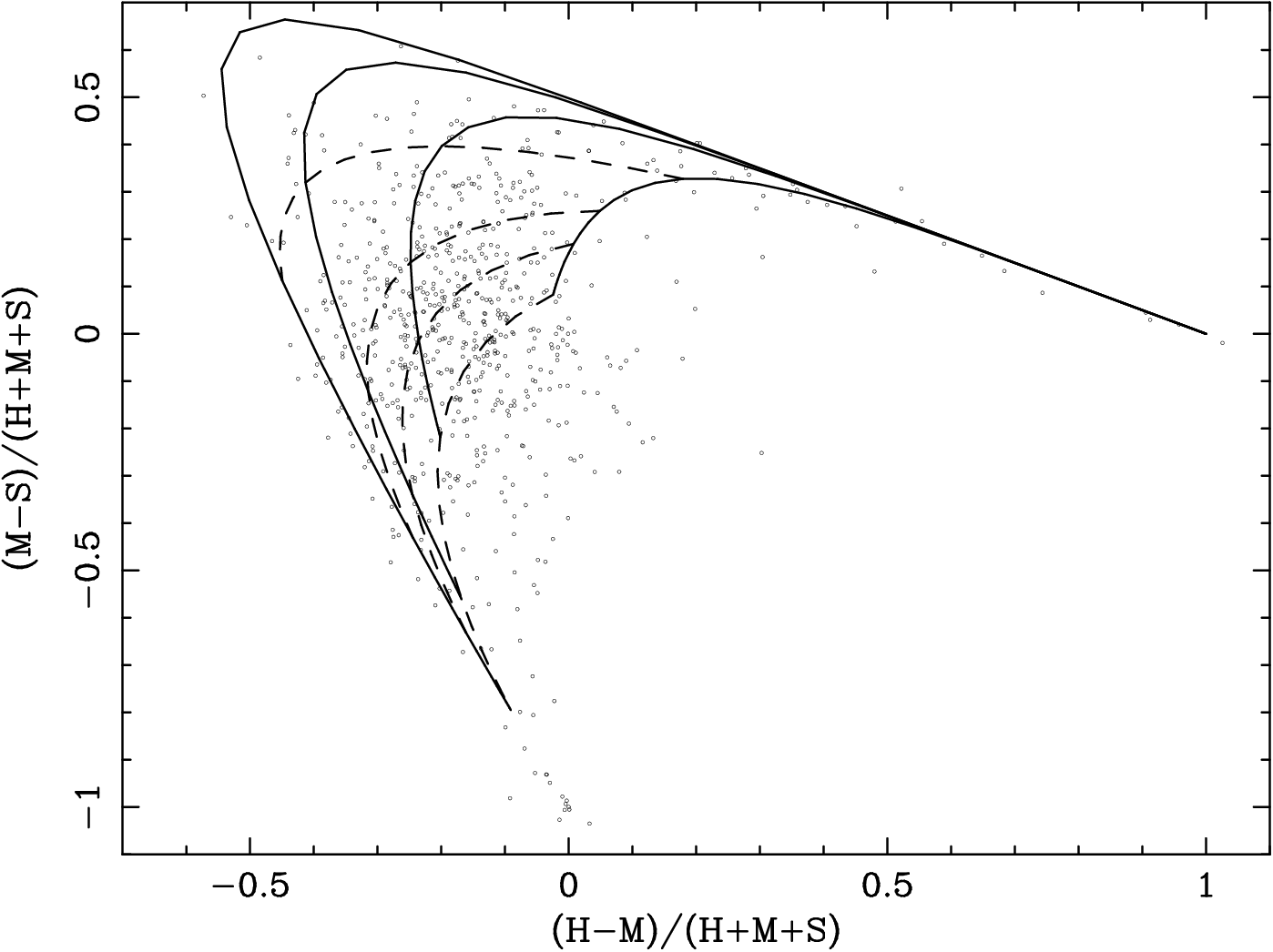}
\figcaption{X-ray colors of sources with luminosities, estimated from observed
counts, in excess of $10^{38}$~\ergl\ and detected counts $>$50. Curves same as Figure~\ref{f:colorU}. Errors omitted for clarity. \label{f:colorAll} }
\end{center}

Highly-absorbed sources lie near (1,0) and
 super-soft sources near (0,-1) in these figures.
The colors of steep power law sources are constrained along a narrow
band extending from the super-soft source location at (0,-1) upwards and
to the left towards $\sim$(-0.6,0.5). The region to the left of this
band is physically inaccessible (curves of constant $\Gamma$
converge rapidly as $\Gamma$ increases as shown in the figures).
According to Prestwich \etal\ (2003), the sources in the region
 centered at about (-0.2,0.0) are predominately low-mass X-ray binaries
 and are common to both spiral and elliptical galaxies.

Visually, the two color distributions appear similar.
Applying a two-dimensional Kolomogorov-Smirnov test
 (Peacock 1983; Fasano \& Franceschini 1987) to the color-color distributions
 of the ULX candidates and of {\sl all} 1900 sources with luminosities above
 $10^{38}$~\ergl\
 resulted in a significance level of $P_{\rm KS}=0.01$.
However, the Kolomogorov-Smirnov test does not account for (statistical)
 uncertainties in the data.
When the test was applied to the restricted set of sources with
 $>$50 detected counts (those depicted in Figures~\ref{f:colorU}
 and~\ref{f:colorAll}), the significance level increased to
 $P_{\rm KS}=0.17$, showing that the distributions are marginally
 consistent with being from the same parent population.

\subsubsection{X-ray Timing} \label{s:RESULTS_timing}

The Kolmogorov-Smirnov probability, $P_{KS}$,
 that a source is constant was computed for the population of sources
 with estimated luminosities exceeding $10^{38}$~\ergl\ and for the
 subset of ULX candidates.
Analysis was limited to those sources with $>$50 detected
 counts as variability is more easily detected at higher signal-to-noise.

\begin{center}
\includegraphics[angle=-90,width=\columnwidth]{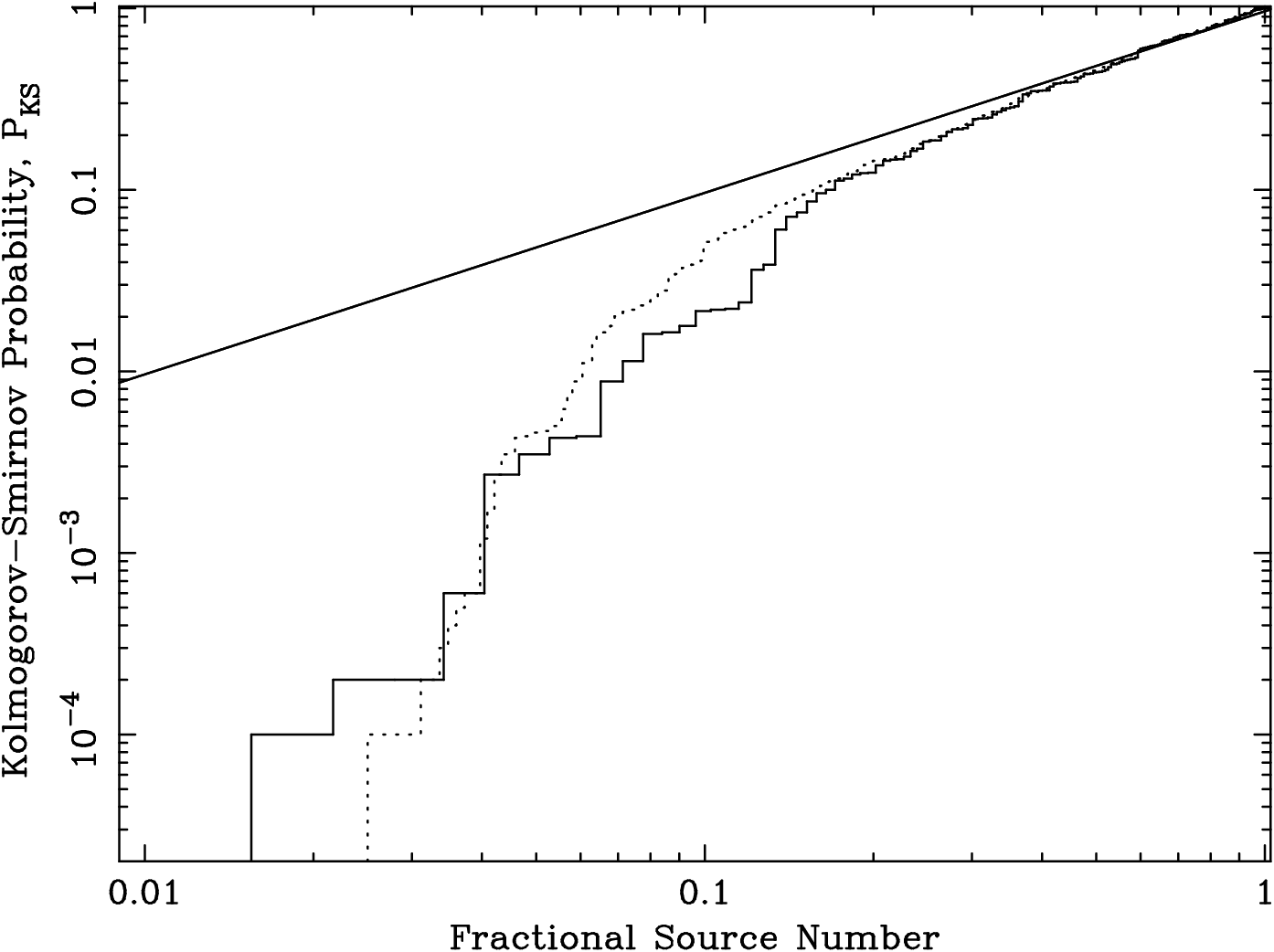}
\figcaption{Results of a Kolmogorov-Smirnov test of the hypothesis that the
distribution of photon arrival times for X-ray sources is equivalent to that
of a source with constant flux. Shown is the value of the significance level,
$P_{KS}$, for each source with estimated luminosity exceeding $10^{38}$~\ergl\
and $>$50 detected counts ({\sl dotted}) and for the ULX
candidates ({\sl solid}) ordered by increasing values of $P_{KS}$.
The line has slope unity.
The abscissa is scaled by the inverse of the number of sources in the two populations ($1/837$ and $1/154$, respectively).
 \label{f:KS1} }
\end{center}

Figure~\ref{f:KS1} displays the distribution of these probabilities for the
 two populations.
The line denotes the linear trend of the data.
For purposes of discussion, we may consider sources with
 $P_{KS}$\LA 0.001 as significantly variable and those with $P_{KS}$\LA 0.04,
 the point where the slope of the curve changes noticably, as likely to
 be variable.
There are 80 sources (9.6\%) with $P_{KS}<0.04$ and 33 (3.9\%) with
 $P_{KS}<0.001$ for the larger sample.
If the ULX candidate population had the same distribution, then 15 and 6 ULXs,
 respectively, would be expected to lie below these values of $P_{KS}$.
For the ULX candidates, there are 22 and 7 sources, or
 14.3\% and 4.5\%, respectively, below these two benchmarks.
This is within statistical errors of the expected values.
Of particular importance is the fact that the photon arrival times for
 86\% of the ULX candidates are consistent with the constant flux hypothesis.

Note that the ULXs tend to be those with higher
 numbers of counts so that variability is more easily discerned with this
 test in this population.
The K-S test is sensitive to long term variability even when
 the data has been finely binned.
This is in contrast to other methods which often have an optimal binning
 typically of the same order as the variability timescale under consideration.
Simulations were run to estimate this sensitivity.
Consider the function $R(1+M{\rm sin}(2 \pi t/P))$, where $R$ is the average
count rate, $M$ is the modulation amplitude, and $P$ is the period.
Events were randomly generated in each timebin with a Poisson distribution
with the mean given by the function.
For 5000 bins and $R=0.02$~c~bin$^{-1}$, the simulation
will generate about 100 c.
For this case and a period of 5000 bins, $M$ needs to be about 0.35
in order for the K-S probability to be 10\%.
In other words, if the modulation is 35\% and the K-S threshold is 10\%
then about half the time the K-S probability will be less than
10\% and the variability ``detected''.
If $P$ is 2500, then the amplitude needs to be 0.60 and for $P$
of 1500 an amplitude of 1.00 only results in a K-S probability of 15\%.
As expected the amplitude scales roughly as $N^{-1/2}$, where $N$ is the number
of counts.

As displayed in Figure~\ref{f:KS1}, the distributions are themselves
 cumulative distribution functions and applying the Kolmogorov-Smirnov test
 to compare these two distributions shows that they are consistent with
 being drawn from the same distribution ($P_{KS}=0.95$).

\subsubsection{X-ray Source Locations} \label{s:RESULTS_positions}

The relative radial position of each X-ray source can be expressed as the
 fraction, $f$, of the deprojected galaxy radius in units of $1/2$ the \elip\
 diameter.
Figure~\ref{f:radial} shows the resulting surface distribution of all 3413
 discrete X-ray sources detected in the survey, of all ULX candidates, of
 ULXs in spiral galaxies, and of ULXs in ellipticals. Recall that sources
 within 5\arcsec of the galaxy centers have been excluded from
 consideration as ULXs.

The distributions can be approximated by a generalized exponential function
 (plus a constant) of the form $A \exp^{-(f/h)^{(1/n)}}$ where
 $h$ is a scale height and $n$ is the index.
The distributions of all detected sources and of all ULX candidates
 can be approximated by the same values of scale height
 ($h=0.06\pm 0.03$) and index ($n=1.59\pm 0.32$, where the quoted errors are
 for the larger population of all detected sources, $\chi^2=105.9$ for 93 dof
 and 25.1 for 17 dof, respectively).
Fits to the smaller subsets of spirals or ellipticals give
 large ranges to the fit parameter values when all parameters
 are allowed to vary. If $n$ is held to the best-fit value of 1.59, then
 the values of the scale heights are within the uncertainties,
 $h=0.06\pm0.04$ ($\chi^2 = 19.9$ for 16 dof)
 for the spirals and $h=0.04\pm0.02$ ($\chi^2 = 21.4$ for 15 dof)
 for ellipticals.

\begin{center}
\includegraphics[angle=-90,width=\columnwidth]{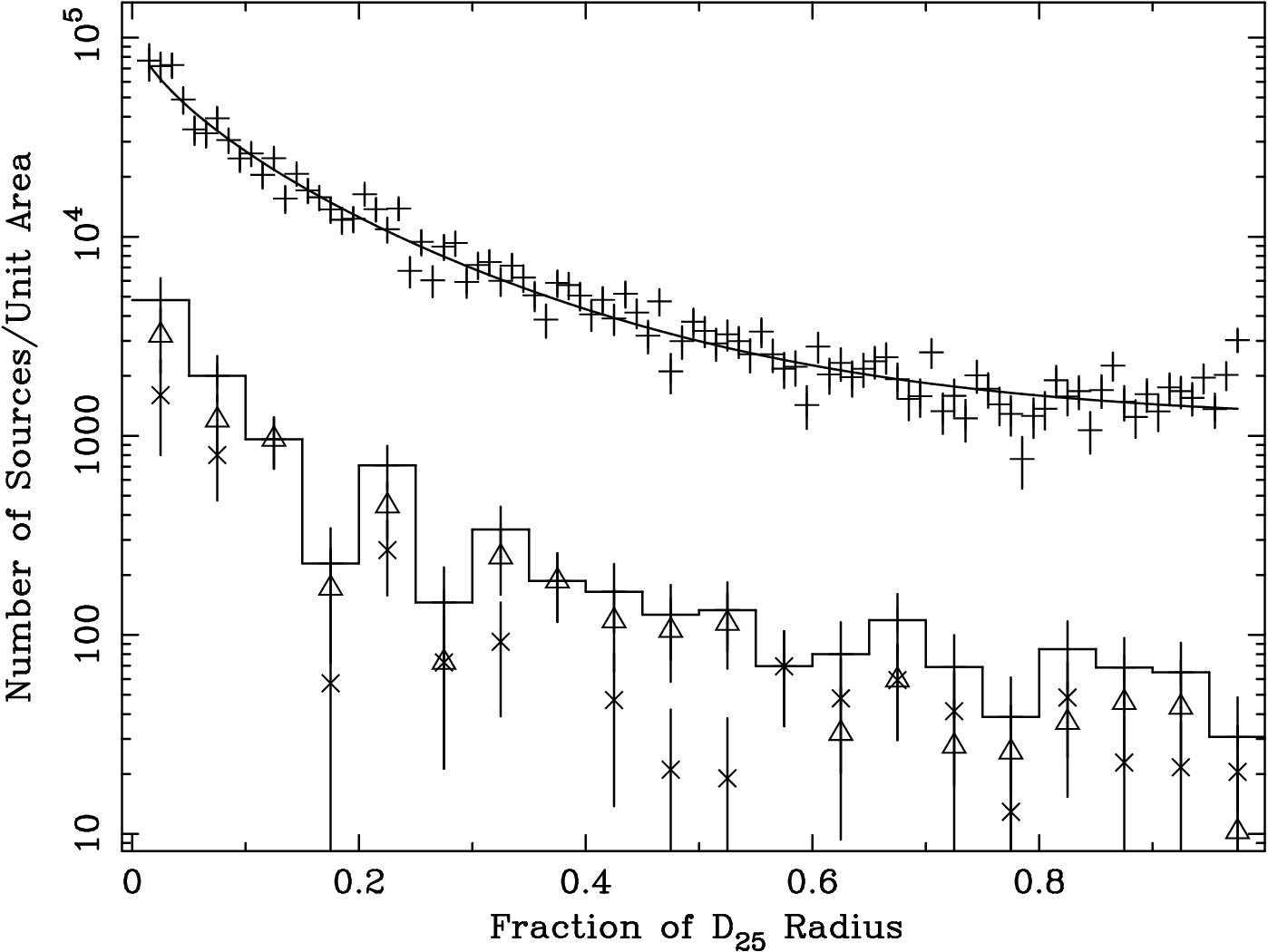}
\figcaption{Surface distribution of all X-ray sources detected in the
sample ({\sl top}) and of the subset of ULX candidates
({\sl bottom}, full curve). ULX candidates in spiral galaxies are marked with
triangles and those in elliptical galaxies with $\times$'s.
The abscissa is the deprojected radial position, $f$, expressed
 as a fraction of the host galaxy's angular radius ($\equiv 0.5$\elip )
 and the ordinate is the number of sources per unit area, $f df$,
 on the range $f$ to $f+df$. The curve drawn through the upper dataset is the
 best-fit generalized exponential function described in the text.
 \label{f:radial} }
\end{center}

Fitting a constant to the surface density distributions
 of the ULX populations in ellipticals and spirals results in an
 acceptable fit for ellipticals but not for spirals.
For ellipticals, the value of the constant is
 $29.1\pm9.1$, $\chi^2 = 23.3$ for 17 dof, on the entire domain.
On the restricted domain, $f>0.1$, where the distribution appears to flatten
 (Figure~\ref{f:radial}), the value is $28.8\pm9.1$, $\chi^2 = 13.9$ for 15 dof.
For spirals, $\chi^2 = 63.7$ for 18 dof and $\chi^2 = 47.4$ for 16 dof
 in these two domains, respectively.
Thus, the ULX candidates
 have the same spatial distribution,
 statistically, as does the larger population of all detected sources but
the spatial distribution of ULX candidates in ellipticals is much flatter
 than for spirals with the exception of a few sources near the centers.

\subsubsection{Estimated Contribution from Background Sources} \label{s:bgsrcs}

The detection limit for discrete sources is about $3 \times 10^{-15}$~\ergcms,
 for the average exposure time of 39.6~ks,
 and the total area of the sky covered by the survey is 0.9~deg$^{-2}$.
Comparing this to the \cxo\ Deep Field (CDF) results of Brandt \etal\ (2001)
 and of Rosati \etal\ (2002), we
 expect about 1100 of the 3400 detected sources are background (consistent
 with the value of the constant term, $1144 \pm 215$,
 in the surface distribution model described in \S \ref{s:RESULTS_positions}).
This corresponds to $\sim$12 background sources per galaxy given the
 average galaxy area is 36.2 min$^2$.

The results for the ULX candidates, however, are based on luminosities and
 not fluxes. Thus, a straightforward comparison to the CDF results
 is not possible.
Instead, in Figure~\ref{f:cdf}, is shown the \lognlogs\ distributions
 of ULX candidates for each galaxy, where $S$ is the observed flux and $N(>S)$
 is the number of sources with flux $>S$ and
 scaled by the area of the galaxy within
 the instrumental field of view in units of deg$^{-2}$.
Also shown is a curve, $N(>S)=16(S/10^{-13})^{-1.7}$,
 representing the high-flux \lognlogs\ results of the \asca\ Large Sky Survey
 (Ueda \etal\ 1998) and Deep Sky Survey (Ogasaka \etal\ 1998) and the
 low-flux results from Brandt \etal\ (2001).
Points near or below this curve are likely
 background objects and not true ULXs.
The number of background sources can be estimated as follows.
If the flux from the weakest ULX candidate in a galaxy (of area $A$~deg$^{2}$)
 is denoted $S$, then we expect a contribution of at most
 $16(S/10^{-13})^{-1.7}A$ to the number of background sources from that
 galaxy field.
If the slope of the galaxy's \lognlogs\ distribution is flatter than
 $S^{-1.7}$, then this is an overestimate.
The sum of such contributions
 from all galaxies gives a conservative estimate of 39 background sources
 among the 154 ULX candidates, or 25\%.

Note that the distribution depicted in Figure~\ref{f:cdf} suggests many ULX
 candidates in elliptical galaxies are more likely to be background sources
 than those in spiral galaxies.
Using the method outlined above yields an estimate of 25 background sources, or 44\%,
 among the elliptical galaxy ULX candidates while only 14 sources (14\%)
 of the ULX candidates in spirals are potentially background objects.
This result is consistent with the
 sample of elliptical galaxies analyzed by Irwin \etal\ (2003).
However, there are
 several ULXs in our sample of
 elliptical galaxies that do not follow this trend.

\begin{center}
\includegraphics[angle=-90,width=\columnwidth]{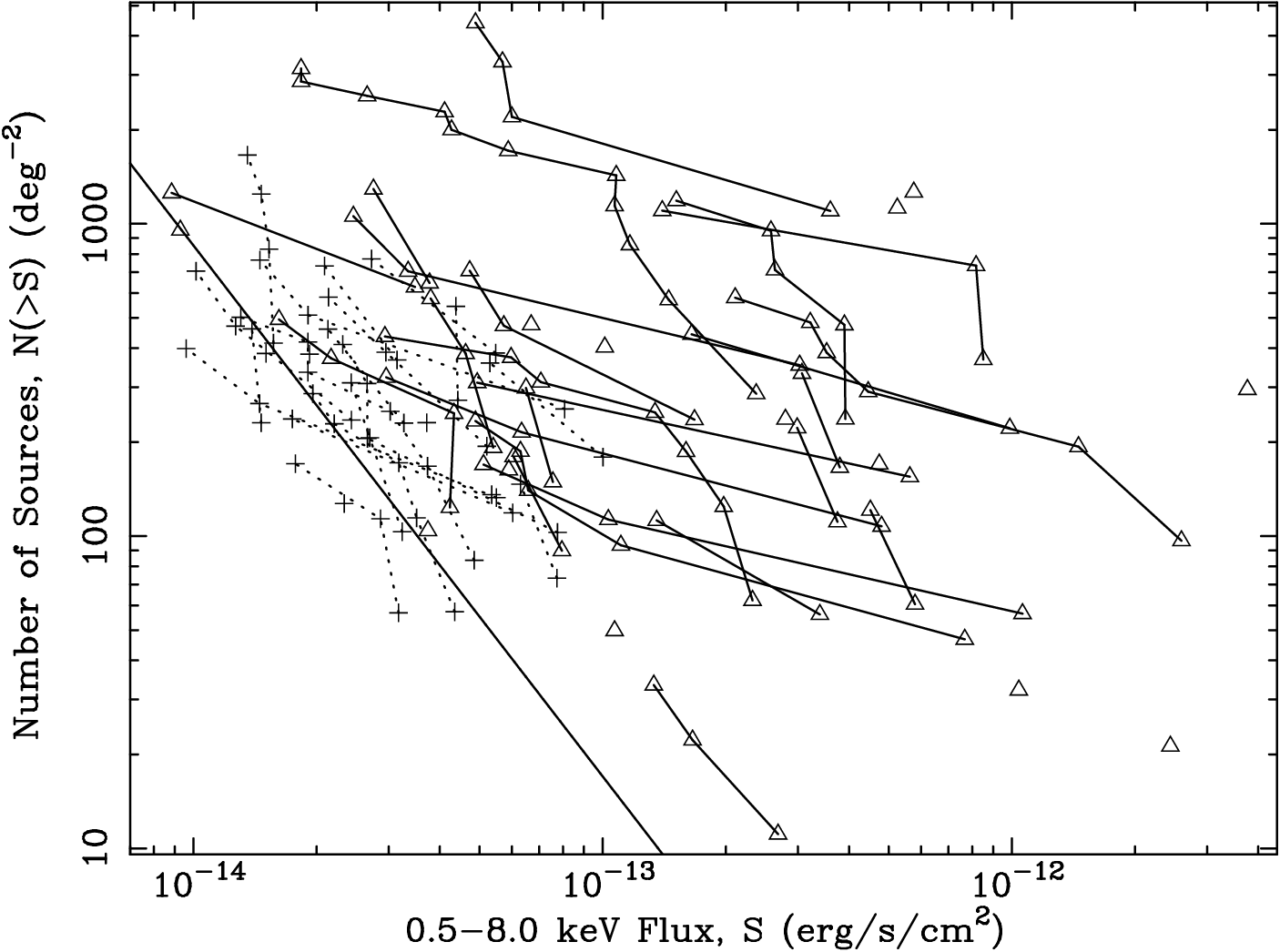}
\figcaption{\lognlogs\ distribution for ULX candidates. The ordinate is
the number of ULX candidates identified in each galaxy above a given flux
scaled by the angular size of that galaxy. Spiral galaxies are denoted
by triangles and solid lines. Ellipticals are denoted by crosses and
dotted lines. The power law
curve denotes an approximate fit to the point-source X-ray background
(e.g., Brandt \etal\ 2001), scaled to our bandpass (0.5--8.0 keV), and
is given by $N(>S)=16(S/10^{-13})^{-1.7}$. Note several galaxies host only
one ULX candidate.
 \label{f:cdf} }
\end{center}

\subsubsection{X-ray Luminosity Function} \label{s:lumfun}

The cumulative luminosity functions (XLFs) of the ULX candidates in elliptical and in
 spiral host galaxies are displayed in Figure~\ref{f:lf}.
There are 57 ULX candidates in the 27 elliptical galaxies and 97 in the 55
 spiral galaxies.
While the number per galaxy are roughly equal, ULXs in spiral galaxies are
 much more luminous. Two-thirds of the ULXs in spiral galaxies are above
 $2\times 10^{39}$~\ergl\ while only $1/3$ of ULXs in ellipticals are this
 luminous.
\begin{center}
\includegraphics[angle=-90,width=\columnwidth]{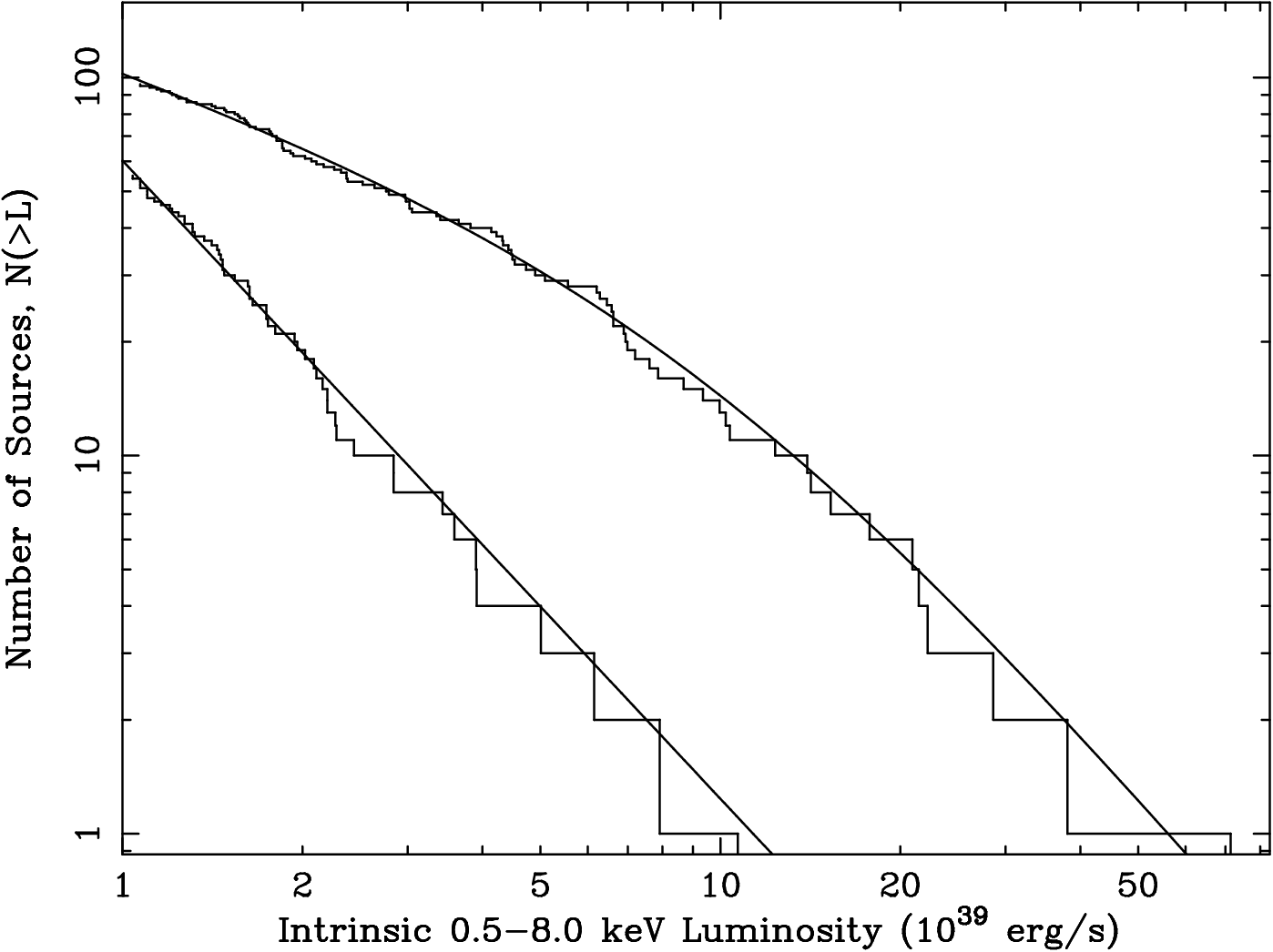}
\figcaption{
Cumulative luminosity function of ULX candidates in spiral (upper curve)
and in elliptical galaxies. A power law with slope -1.72$\pm$0.14 describes
the luminosity function for ellipticals. A broken power law model
is shown for spirals.
(A power law with an exponential cut-off also provides an acceptable fit for
spirals though a single power law does not.)
The slope is -0.56$\pm$0.05 at low luminosity and the break
luminosity is $\sim$$10 \times 10^{39}$~\ergl.
 \label{f:lf} }
\end{center}
Using a maximum-likelihood statistic,
 a power law adequately describes the luminosity function for ellipticals
 with a slope -1.72$\pm$0.14. More complex models do not improve the fit
 significantly.
A broken power law
 (change in statistic $\Delta C=26.7$ compared to a simple power law)
 or a power law with an exponential cut-off ($\Delta C=23.8$) is needed
 to describe the luminosity function for spirals.
The slope of the broken power law
 is -0.56$\pm$0.05 at low luminosity, the break luminosity is
 9.6$\pm$0.6 in units of $10^{39}$~\ergl, and the slope at high luminosity is
 -1.88$\pm$0.30.
The slope of the cut-off power law model is
 -0.64$\pm$0.09 with a cut-off luminosity of
 $(19.4\pm 8.6) \times 10^{39}$~\ergl.

\subsection{ULXs and their Host Galaxies} \label{s:hosts}

Several global properties of the host galaxies are available as described
 in \S~\ref{s:METHODS_galProps}.
Linear correlation coefficients were computed for various paired
 combinations of the following quantitites compiled for each galaxy:
 number of ULXs ($N_{\rm ULX}$);
 total X-ray luminosity from ULXs ($L_{\rm ULX}$)
 average ULX luminosity ($L_{\rm ULX}/N_{\rm ULX}$);
 blue luminosity (\lb ), far-infrared luminosity (\lfir ),
 nearest-neighbor distance ($\theta_p$), and \elip\ area.
Linear correlation coefficients for the
 most significant correlations are listed in Table~3.
In addition to the entire sample of galaxies, coefficients were computed
 for the subsets of elliptical and of spiral galaxies and for the subsample of
 \cxo\ galaxies with the same distribution of absolute $B$ magnitudes as the
 Tully catalogue of galaxies (see \S~\ref{s:RESULTS_sample}).

The strongest correlations for elliptical galaxies are between ULX properties and
 the host galaxy's \lb, while those for spirals are with \lfir\ and a
 marginally-significant correlation with $\theta_p$.
There are no significant correlations for the subsample of elliptical galaxies.
The subsample of spiral galaxies again correlate with \lfir\ but also with \lb.
The subsample of galaxies was designed to minimize some of the biases in the
full sample. Although the subsample contains only 34 galaxies,
the significance of
the correlations betwen ULX properties and \lfir\ for spirals remains strong
enough to show this result is robust.

While there are $\sim$2 ULX candidates per galaxy for both
 ellipticals and spirals (\S \ref{s:lumfun}), the number per unit
 ($10^{42}$~\ergl ) blue luminosity for ellipticals is only 0.11$\pm$0.02
 compared to 0.30$\pm$0.11 for spirals.
The total ULX luminosity, $L_{\rm ULX}$, per unit \lb\ is
 0.19$\pm$0.05 for ellipticals and 1.63$\pm$0.88 for spirals.
Therefore, the pairs of parameters were again examined after first
 normalizing to unit \lb.
No significant changes in the correlation coefficients occurred though the
correlations between ULX properties and \lfir\ increased for spiral galaxies.

The correlation between the number of ULX candidates and \lfir\ is
 displayed in Figure~\ref{f:NulxVsLfir} for the spirals and ellipticals.
For spirals, the linear relation between the two is
 $N_{\rm ULX}$$=$$(0.022 \pm 0.01)L_{\rm FIR}/10^{42}+(0.64\pm0.3)$.
Note that the highest \lfir\ bin is dominated by the Antennae pair of galaxies
 with 13 ULX candidates.
On the other hand, the spiral galaxies with \lfir \LA $4\times 10^{42}$~\ergl\
 are the dwarf irregulars ({\sl cf}. Figure~\ref{f:BvF}).
For ellipticals, the strongest correlation is between the number of ULXs
 and {\sl blue} luminosity and is given by
 $N_{\rm ULX}$$=$$(0.095 \pm 0.05)L_{\rm B}/10^{42}+(-0.11\pm0.7)$.

\begin{center}
\includegraphics[angle=-90,width=\columnwidth]{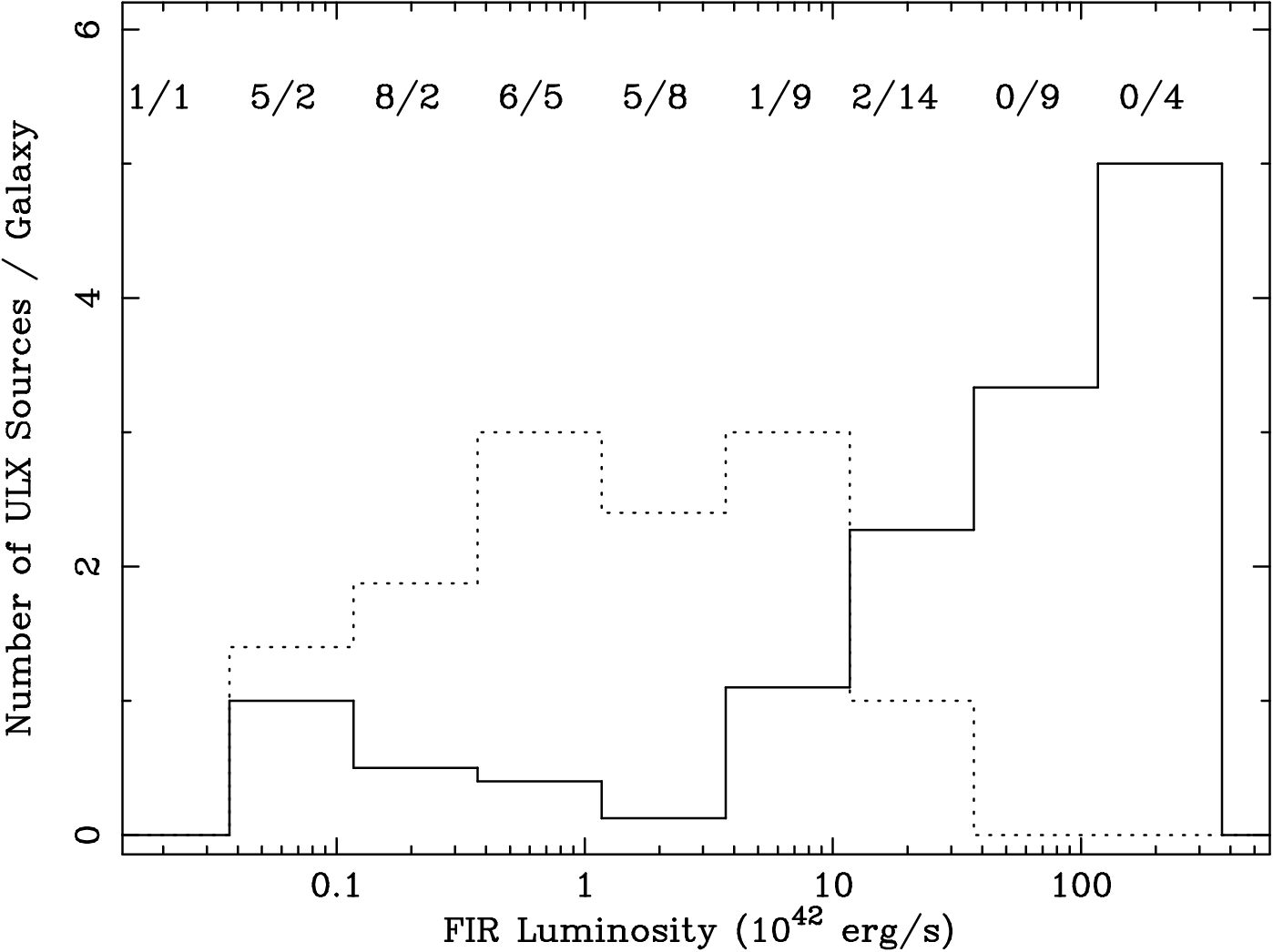}
\figcaption{Number of ULX candidates per galaxy is shown against the host
galaxy's far-infrared luminosity. For spiral galaxies ({\sl solid} histogram),
the correlation is significant (Pearson's linear correlation
coefficient $r=0.88$) but not for ellipticals ({\sl dotted}, $r=0.15$).
The numbers of elliptical/spiral galaxies within each \lfir\ bin are displayed
across the top of the figure.
 \label{f:NulxVsLfir} }
\end{center}

\section{DISCUSSION} \label{s:DISCUSSION}

We have presented the spectrophotometric
 X-ray properties of a population of 154 ULX candidates
 taken from a sample of 82 galaxies observed with \cxo.
We have shown that ULXs are more numerous and luminous
 in galaxies with indications of
 recent star formation as measured by their far-infrared luminosities and by
 their morphologies (Hubble type and, less significantly, nearest neighbor
 distance).
We have also compared the ULX candidates with the less-luminous X-ray sources
 in the same galaxy fields and found their X-ray properties statistically
 indistinguishable.

\subsection{ULXs and Less-Luminous Sources}

The similarity between ULX candidates and the less-luminous population
 exists among all observable X-ray properties: spectral shapes, colors,
 time series, and (radial) positions within their host galaxies.
This result has two important implications.
First, as the less-luminous population is a composite of
 low- and high-mass X-ray binaries and supernova remnants (to our formal detection
 limit but also must include some foreground objects such as cataclysmic variables and stellar coronae
 as well as background AGNs); our results imply the ULX candidates are also
 a heterogeneous group of objects.
Secondly,
the similar distributions of physical characteristics suggests, as the
 simplest explanation, that
 ULXs may be the high-luminosity end of a continuous distribution of
 supernova remnants and accreting systems such as
    X-ray binaries.

These conclusions do not diminish
 the fact that ULXs retain their distinction as
 the most luminous non-nuclear X-ray sources.
How are such high luminosities achieved?
Contemporary explanations for the ULX phenonema fall into three broad
 catagories; massive accreting objects, anisotropically-emitting sources,
 and supernovae.
As part of their appeal, each of these models effectively circumvent the
 benchmark Eddington limit luminosity,
 $L$$\sim$$1.4\times10^{38} (\mbox{M/\msun})$~\ergl,
 that expresses the balance between the
 inward force of accreting material against the outward radiation force.
X-ray binaries with compact object masses $>$100~\msun\ raise the
 Eddington limit to $>$$10^{40}$~\ergl.
Anisotropically-emitting X-ray binaries avoid the Eddington limit
 in two ways: The emission may be in a direction other than opposing
 accretion and, secondly, the actual luminosity is
 less than the apparent (assumed isotropic) luminosity by a factor
 $b=\Omega/4\pi$ where $\Omega$ is the solid angle of emission.
For accreting sources, it is also necessary to attain an accretion rate,
 \mdot \GA $10^{-6}$$b$~\msun ~yr$^{-1}$, as the apparent luminosity scales as
 $\sim$$6 \times 10^{45} \dot{M}/b$~\ergl\ [and so the slim-disk scenario
 that applies at high accretion rates
 (Abramowicz \etal\ 1988; Watarai \etal\ 2001; Ebisawa \etal\ 2003) may be
 favored over standard thin-disk models].
The third catagory, supernovae, are, of course, not Eddington
 limited in the first place.
Each of these scenarios predict different observable X-ray properties
 which can be compared to our results.

\subsubsection{Intermediate-Mass Black Holes}

Colbert \& Mushotzky (1999; see also Makishima \etal\ 2000;
 Colbert \& Ptak 2002) suggest that ULXs are accreting black holes (BHs) with
 masses intermediate between stellar mass (\LA 20 \msun) and
 supermassive objects (active galactic nuclei, $10^6$ to $10^9$ \msun).
This is perhaps the most intriguing scenario because it suggests a possible
 observable link between stellar collapse and the formation of active
 galactic nuclei (see van~der~Marel 2004; Miller \& Colbert 2003 for reviews).
Fryer \& Kalogera (2001) show, however, that single stars in the current
 epoch rarely form BHs with masses \GA 15~\msun.
First-generation, zero-metallicity (Population III) stars, on the other hand,
 were on average
 more massive than today's population, would have retained more mass as
 they evolved because of inefficient wind mass loss, and may have
 left more massive BH remnants.
Intermediate mass black holes (IMBHs) may also form in dense
 young star clusters through runaway merging of massive stars
 (Ebisuzaki \etal\ 2001; Portegies~Zwart \& McMillian 2002;
 Portegies~Zwart \etal\ 2004) or through
 the gradual accural of compact stellar remnants
 onto a seed black hole over the lifetime of the
 cluster (Miller \& Hamilton 2002).

In analogy to stellar-mass black hole X-ray binaries
 (e.g., McClintock \& Remillard 2003), the IMBH X-ray spectrum
 should be a power law with a soft thermal (blackbody) component
 with a characteristic temperature scaling as the -1/4th power of
 the compact object mass (e.g., Makishima \etal\ 2000).
Those ULX candidates with blackbody disk spectral shapes typically
 have temperatures of order 1~keV (with two exceptions which lie near
 $(0,-1)$ in the color-color diagram, Figure~\ref{f:colorU}, i.e., they are
 supersoft sources with no additional power law component).
Thus, we find no prominent component at $kT$$\sim$0.1--0.2~keV in our sample.
A few ULX candidates with soft accretion disk components have been
 reported (e.g., Miller \etal\ 2003; Roberts \& Colbert 2003).

To further the analogy to stellar-mass BHs, IMBHs can also be expected
 to display soft/hard spectral transitions.
Furthermore,
 transient behavior is predicted by Kalogera \etal\ (2004) for accreting
 IMBH systems due to a thermal-viscous disk instability.
Neither of these
 transitions are accessible in the short observations typical
 of our sample but have been reported for a few well-monitored ULXs
 (e.g., La~Parola \etal\ 2001; Kubota \etal\ 2001).

IMBHs formed from individual Population~III stars are expected to be
 distributed throughout the host galaxy and to be  present in all galaxy types
 (Madau \& Rees 2001).
IMBHs formed in dense stellar environments should remain associated with
 globular clusters.
They should occur frequently in elliptical galaxies as these
 galaxies have large numbers of globular clusters.
Furthermore, the ULX spatial distribution should follow that of the halo stars
 in contrast to the distribution of weaker sources that follow the distribution
 of disk and bulge stars.
These trends are difficult to discern in the current sample.
The ULXs are distributed in the same way as the weaker sources (though the
 distribution in ellipticals is arguably flatter, \S \ref{s:RESULTS_positions})
 and
the number of ULXs per galaxy is no higher in cluster-rich
 ellipticals compared to spirals.
Note that an IMBH accretor need not be invoked
 for the majority of ULXs in elliptical galaxies where $2/3$ of the
 ULXs (\S~\ref{s:lumfun})are less luminous than $2 \times 10^{39}$~\ergl\
 and, according to Irwin, Bregman, \& Athey (2004), those more luminous than
 $2 \times 10^{39}$~\ergl\ may all be background objects.

\subsubsection{Anisotropic X-ray Binaries}

There are several beaming models:
One model (King \etal\ 2001)
 assumes a phase of super-Eddington accretion is present, perhaps due to
 thermal-timescale mass transfer, that forms a
 thick disk with a central funnel that results in mildly-beamed emission
 (see also Madau 1988).
Another model envisions the formation of a jet with an enhancement of the
 X-radiation via Compton scattering of photons (from the accretion disk or
 from a high-mass companion) by relativistic electrons in the jet
 (Georganopoulos \etal\ 2002).
A third model envisions direct, relativistic Doppler boosted,
 synchrotron emission from the (face-on) jet itself
 (Fabrika \& Mescheryakov 2001).

The spectral signature of the beaming model has not been specified.
Compton scattering from jets and direct synchrotron emission should display
 a characteristic steep power law spectra.
If ULX emission is beamed and weaker sources are not, then one would expect
 the two populations to have different spectral indices.
While the spectral shape of the majority of the ULX candidates are consistent
 with power laws, their spectral indices do not differ from the less luminous
 population.
Large-amplitude variability is also expected from relativistic jets.
The high mass transfer rates implied for the mildly beamed sources
 (King \etal\ 2001; Kalogera \etal\ 2004)  result in
 stable disks and persistent X-ray emission.
There are no preferred locations or galaxy type for beaming models.
However, thermal-timescale mass transfer occurs when the
 main-sequence mass-donor
 (companion) star is more massive than the compact accretor.
This requires an early-type companion of moderate mass, \GA 5~\msun, and hence
 such systems will occur rarely in ellipticals.

\subsubsection{Energetic Supernovae}

A third scenario is an unusually energetic supernova, or hypernova (Wang 1999).
Supernovae exploding in high-density environments have
 been hypothesized to reach X-ray luminosities sufficient to power the
 broad-line regions of active galactic nuclei
 (Terlevich, Melnick, \& Moles 1987; Terlevich \etal\ 1992).
Supernovae have been observed with luminosities as high as
 $\sim$$10^{40}$~\ergl\ months to years after their discovery
 (Schlegel 1995; Immler \& Lewin 2003).
For example, SN~1986J in NGC~891 has a luminosity of
 $\sim$$6\times 10^{39}$~\ergl\ accoding to our calculations (Table~2).

Supernovae display the distinct thermal emission-line X-ray spectrum from
 shock-heated plasma; often with a harder bremsstrahlung component from a
 forward shock.
This spectrum is easily distinguished from a power law shape.
Only $\sim$7\% of ULX candidates {\sl require} a thermal model to reproduce their
 spectra.
As shown in \S~\ref{s:RESULTS_spectra}, about 15\% of the ULX
 candidates with acceptable power law fits have steep slopes, $\Gamma$\GA 3,
 suggesting a thermal spectrum.
Thus, at least 22\% of the ULX candidates are likely thermal sources.
The actual number could be higher since a
 thermal model does provide a formally-acceptable fit to 78\% of the ULX candidates.
Nevertheless, it is unlikely that the majority of ULX candidates are thermal
 sources since the average power law model slope of 1.8
 (or the average thermal model temperature of 18.4~keV)
 is much more descriptive of non-thermal sources.

Supernovae should be steady sources on short timescales and decline
 slowly over periods of months to years.
Again, while results of variability tests are consistent with the majority of
 ULX candidates being steady sources, individual observations are too short to
 discern long-term trends and some low-level variability may be hidden in the
 low-count light curves.
Variability has been reported for several ULX candidates based on multi-epoch
 monitoring (e.g., Fabbiano \etal\ 2003; Roberts \etal\ 2004, La~Parola \etal\ 2001).
In combination with detailed spectral modeling and radio wavelength imaging,
 the number of energetic supernovae contributing to the ULX population should soon
 be much better constrained.
Core collapse supernovae have massive star progenitors
 consistent with the association of ULXs with late-type star-forming galaxies.

\subsection{ULXs and their Host Galaxies}

It has long been known (David, Jones, \& Forman 1992)
 and recently confirmed
  (Ranalli, Comastri, \& Setti 2002; Grimm, Gilfanov, \& Sunyaev 2003)
 that the total X-ray luminosity of a galaxy
 correlates with recent star formation rate as measured
 by its far-infrared luminosity, \lfir.
We have shown that this trend extends to the numbers and total luminosities
 of ULXs found in such galaxies.
This result was anticipated by the large numbers of ULXs in some FIR-bright
 galaxies (e.g., NGC~4038/4039, Zezas \etal\ 2002;
 NGC~3256 Lira \etal\ 2002; NGC~4485/4490, Roberts \etal\ 2002)
 and for a small sample of galaxies by Grimm, Gilfanov, \& Sunyaev (2003).
A similar conclusion can be indirectly deduced
 through studies of the cumulative X-ray
 luminosity functions of nearby galaxies:
The flat slopes of the XLFs implies that the
 most luminous sources dominate the total X-ray luminosity
 of spiral and starburst galaxies
 (e.g., Kilgard \etal\ 2002; Grimm \etal\ 2003; Colbert \etal\ 2004).

Two of the theoretical explanations for ULXs discussed in the previous
 section predict an association of ULXs with recent star formation.
These are the mildly-beamed thermal-timescale mass transfer X-ray binaries
 described by King \etal\ (2001) and X-ray bright supernovae.
While a case might be envisioned for IMBHs formed in prompt collapse of the
 cores of young super-star clusters,
 studies of the Antennae by Zezas \etal\ (2002, see also Clark \etal\ 2003)
 and of three other starburst galaxies by Kaaret \etal\ (2003)
 determined ULXs are often near but
 not spatially coincident with these clusters
 as would be expected if they are IMBH binaries.
Recent numerical simulations (Portegies Zwart \etal\ 2004) indicate the
 formation frequency of IMBHs in super-star clusters depends sensitively
 on cluster initial conditions and dynamical friction timescale and thus
 does not simply scale with the host galaxy's (global) star formation rate.
As only a small fraction of ULX candidates in our survey display the requisite
 thermal emission-line spectrum characteristic of supernovae, there remains
 only the model of King \etal\ (2001) as a reasonable representation of the
 {\em majority} of ULXs.

Does the model of King \etal\ (2001) provide
 a viable explanation for the results of the present survey?
The model is essentially that of a high-mass X-ray binary (HMXRB)
 undergoing an episode of thermal-timescale mass transfer.
Such systems arise when the donor star envelope is in a radiative phase either
 when it is more massive than its companion or first fills its Roche lobe while
 expanding to the red giant stage.
Particulars of the mass transfer mechanism, possibilities of common envelope
 formation, and details of beaming and resulting luminosities
 are beyond the scope of the present work.
However, the question remains
 if enough systems of this type occur to account for the
 numbers of ULXs reported here.

Assuming $\sim$2\% of OB stars form HMXRBs (e.g. Dalton \& Sarazin 1995,
 Helfand \& Moran 2001),
 that the thermal-timescale phase always occurs and lasts $\sim$$10^5$ yr
 (King \etal\ 2001) or 1\%
 of the lifetime of a star (10--40 Myr for stars of initial mass 8--20~\msun),
 and a beaming factor $b=0.1$; then at least
 $\sim$$5\times 10^4$ OB stars per ULX are required.
From results of evolutionary synthesis models,
 Leitherer \& Heckman (1995) find
 this many O~stars are formed for a star formation rate (SFR)
 of $\sim$2~\msun\ yr$^{-1}$ over a period of $>$10~Myr.
This result is for a particular choice of initial mass function,
 metallicity, upper mass cutoff, and star formation time-scale
 representative of conditions found in infrared-luminous starburst
 galaxies.
The relation between \lfir\ and SFR deduced by Leitherer \& Heckman (1995)
is\footnote{\lfir\ in this expression is the 8--1000~$\mu$m luminosity which
is about 50--100\% larger than the 42.5--122.5~$\mu$m luminosities listed in
Table~2 assuming a warm dust temperature.}
 SFR$=$$0.045(L_{\rm FIR}/10^{42}$)~\msun\ yr$^{-1}$ (see also Kennicutt 1998).
These values predict $\sim$2 to 5 ULXs per \lfir $=$$10^{44}$~\ergl,
 though the uncertainties are large.
From \S \ref{s:hosts}, the number of ULXs per (spiral) galaxy is
 $\sim$1 for galaxies with \lfir $\sim$$10^{43}$~\ergl\ rising to
 $\sim$5 for \lfir \GA $10^{44}$~\ergl.
Thus, the number of ULXs in high \lfir\ galaxies follows the trend expected for
 a short-lived high stellar mass origin.
Leitherer \& Heckman (1995) also predict a peak massive-star
 supernova rate of 0.02~yr$^{-1}$
 at a SFR of 1~\msun\ yr$^{-1}$ which will also contribute to the ULX population
 of active star-forming galaxies in proportion to \lfir.

For lower \lfir, and particularly for elliptical galaxies, the number of ULXs
 is larger than this simple estimate predicts.
Being dominated by the elliptical galaxies, ULXs in this group are
 correlated most strongly with the host galaxy's \lb.
The number of ULX candidates per elliptical galaxy scales roughly as
 $N_{\rm ULX}$$\sim$$0.1(L_{\rm B}/10^{42})$ (\S~\ref{s:hosts}).
It is thus natural to relate the lower-luminosity
 ULXs to the long-lived population of stars,
 namely, to the low-mass X-ray binaries (LMXRBs).
ULX candidates in weak \lfir\ galaxies are also less-luminous than
 those in active star-forming galaxies as
 the average ULX luminosity also correlates with \lfir\ (\S~\ref{s:hosts}).
The XLFs depicted in Figure~\ref{f:lf} show that the ULXs in ellipticals are
 less-luminous than those in spirals.
Irwin, Athey, \& Bregman (2003) and Irwin, Bregman, \& Athey (2004)
 show further that all ULX candidates in
 elliptical galaxies more luminous than $2\times 10^{39}$~\ergl\ are likely
 background sources.

LMXRBs with stellar-mass ($M$$\sim$10--15~\msun ) BH accretors can reach
 luminosities of $\sim$$2\times 10^{39}$~\ergl\ without violating the
 Eddington limit.
The difficulty is achieving accretion rates of up to
 $\sim$$10^{-6}$~\msun ~yr$^{-1}$.
Although such high rates are seen in the soft X-ray transients
 (King 2002; Terashima \& Wilson 2004), persistent sources within
 our Galaxy do not achieve this rate.
A typical Galactic LMXRB radiating at a few $10^{37}$~\ergl\ has an
 accretion rate of $\sim$$10^{-8}$~\msun ~yr$^{-1}$ and
 can maintain this luminosity for
 $M/\dot{M}$$\sim$$10^8$~yr for a 1~\msun\ mass-donor star.
If all accretion rates are equally likely, then
 the probability of observing a ULX in a single observation
 would be inversely proportional to the lifetime at that accretion rate
 (e.g., Wu 2001).
For this assumption, a luminosity of $\sim$$10^{37}$~\ergl\ is expected to be
 100 times more common than $\sim$$10^{39}$~\ergl.
This is consistent with what is inferred from the
 ULX cumulative luminosity function for
 ellipticals, Figure~\ref{f:lf}:
There is one ULX per 100 X-ray sources above 6.3$\times$$10^{37}$~\ergl.
Of order 100 X-ray sources per \lb $=$$10^{43}$~\ergl\ are routinely detected
 above a few $10^{37}$~\ergl\
 in elliptical galaxies (e.g., NGC~4697, Sarazin \etal\ 2001) and in the
 bulges of spiral galaxies (e.g., NGC~3031, Swartz \etal\ 2003).
The average blue luminosity of the elliptical galaxies in our sample is
 $(18\pm 3) \times 10^{42}$ (Figure~\ref{f:BvF}).
Thus, the observed rate of $\sim$2 ULX candidates
 per elliptical galaxy (or 1 ULX candidate per \lb $=$$10^{43}$~\ergl)
 is consistent with a LMXRB origin.
The actual number is somewhat less since nearly 50\% of the ULX candidates in
 elliptical galaxies are potentially background sources (\S~\ref{s:bgsrcs})
 but this line of reasoning suggests that the dominant factor determining
 the slope of the cumulative X-ray luminosity function for ellipticals is
 simply the lifetimes of the LMXRBs.

Low-mass XRBs also exist in spiral galaxies.
However,
inspection of Figure~\ref{f:BvF} shows that, for spirals, \lb\ is more a measure
 of the young stellar content than of the total number of stars
 (or galaxy mass).
Thus, while the contribution of LMXRBs to the ULX population is expected to
 scale with galaxy mass,
 it may not vary linearly with \lb\ for spirals in the present sample.
Colbert \etal\ (2004) used $B$ and $K_s$ luminosities as a measure of galaxy
 mass to conclude that more than 20\% of ULXs in spiral galaxies
 originate from the older stellar population.

Arguments for contributions from the young and the old
 stellar populations to the total X-ray luminosity of nearby
 galaxies have been made since the \ein\ Observatory era
 (see the review by Fabbiano 1989).
More recently, Colbert \etal\ (2004) show that X-ray point source populations
 can also be described as a superposition of contributions from these two
 stellar populations.
Here it has been shown that the ULX candidate population also
 displays this dichotomy.
Considering ULXs as a subset of all X-ray sources
 or, equivalently, as a subset of the total X-ray luminosity of a galaxy,
 this result supports the idea that ULXs are the high-luminosity end
 of a distribution of HMXRBs and supernovae and of LMXRBs.

\subsection{Summary}

X-ray properties and celestial positions of a sample of 154 ULX candidates
 have been tabulated.
The X-ray properties alone are unable to discriminate between ULXs and the
 less-lumious population (to a formal completeness limit of
 $\sim$$4\times 10^{38}$~\ergl).
This could perhaps be foreseen since, for example,
 the power law indices of accreting X-ray sources
 ranging from stellar-mass neutron stars and black holes to supermassive active
 galactic nuclei are, on average, comparable to the mean power law index,
 $\Gamma$$=$1.74$\pm$0.03, obtained here for ULX candidates.
Based on spectral shape, namely a steep power law index or, equivalently,
 a cool plasma temperature, about 20\% of the ULX candidates are
 potentially supernovae but this group also includes a few variable
 and/or supersoft sources.
The supersoft (or `quasisoft') spectra may be a signature of IMBHs as
 these objects are expected to have a soft blackbody disk component.
Most of the sources studied here, however, can be described by a simple
 power law and do not require an additional (soft) thermal component.
Future spectrophotometric X-ray observations will help distinguish between
 transient XRBs and slowly-evolving young supernovae but optical and radio
 data (and rapidly-sampled X-ray timing data) will be needed to discern, e.g.,
 beamed sources from IMBHs.
The precise source locations provided by \cxo\ should be a tremendous aid
 to such followup observations.

The number of ULX candidates per galaxy is roughly the same for spirals and
 ellipticals but
 the luminosities of the ULXs are $\sim$10 times higher in spirals
 compared to ellipticals and the ULX candidate population in ellipticals 
 is severely compromised by background sources (estimated to be 44\% in the
 present sample compared to only 14\% for spirals).
There is also a strong correlation between the number and average luminosity of
 ULXs in spirals and their host galaxy's far-infrared luminosity;
 and a weaker (anti-)correlation with nearest-neighbor distance.

The X-ray data do not favor any particular model for ULX phenomena.
The similarity to properties of weaker sources strongly suggests ULXs are
 not a distinct class but composed of a heterogeneous mixture as are the
 weaker sources; in which case
 all models remain viable on a case by case basis.
The association with recent star formation, and the significant lack of
 bright ULXs in elliptical galaxies, does make a strong case for young,
 short-lived, systems such as HMXRBs as the dominant component of the more
 luminous ULX population.
On the other hand, the population of ULXs in ellipticals and
 at least some of the ULXs in spirals can be explained as the
 high-luminosity end of the LMXRB population.

The intrinsic luminosities of the ULX candidates in the sample range up to
 $\sim$$10 \times 10^{39}$~\ergl\ in the 0.5--8.0~keV band, the point where
 the luminosity function rolls over, with a few objects radiating at
 even higher luminosities.
The Eddington luminosity limit thus requires accretor masses of order
 80~\msun\ and possibly even higher.
Stellar-mass BHs can form with masses up to $\sim$20~\msun\
 (Fryer \& Kalogera 2001).
If a companion of high initial mass can add some
 10--20~\msun\ during a phase of super-Eddington accretion and the system
 is observed near the endpoint of this phase, then the luminosity of such
 a high mass X-ray binary could readily be of order $10 \times 10^{39}$~\ergl\
 if the Eddington limit is exceeded by modest amounts and a high accretion
 rate can be acheived.
While it may still be difficult
 to attain the highest luminosities inferred in the current study,
 this need only occur rarely to account for the handful of very
 luminous ULXs (among the thousands of sources detected).

\acknowledgements

We thank Roberto Soria for discussions on the nature of ULXs
and Jimmy Irwin for insightful conversations on the role of background
sources in the fields of elliptical galaxies.
We thank the referee for a thorough critique of the manuscript and an
alternative perspective on many key points.
This research has made use of the NASA/IPAC Extragalactic Database (NED) which
 is operated by the Jet Propulsion Laboratory, California Institute of
 Technology, under contract with NASA;
of Second Palomar Observatory Sky Survey
 (POSS-II) images made by the California Institute of Technology with funds
 from the NSF, NASA, the National Geographic Society, the Sloan Foundation, the
 Samuel Oschin Foundation, and the Eastman Kodak Corporation;
of Digitized Sky
 Survey images produced at the Space Telescope Science Institute under U.S.
 Government grant NAG W-2166;
of the \cxo\ X-ray Center ObsID-based literature
 search engine, http://asc.harvard.edu/cda/bib.html;
and of software
 obtained from the High Energy Astrophysics Science Archive Research Center (HEASARC),
 provided by NASA's Goddard Space Flight Center.
Support for this research was provided in part by
NASA/\cha\ grant AR2-3008X.

\clearpage
\begin{deluxetable}{llrrrrrrrlrl}
\tablecolumns{11}
\tablewidth{0pt}
\tabletypesize{\scriptsize}
\tablenum{1} 
\tablecaption{Properties of Sample Galaxies \& X-ray Observation Summary}
\tablehead{ \colhead{Galaxy} & \colhead{Type} &  \colhead{Distance$^a$} & \colhead{$L_{\rm B}/10^{42}$} & \colhead{$L_{\rm FIR}/10^{42}$} & \colhead{$\theta_p^b$}  & \colhead{ObsID} & \colhead{Date} & \colhead{$t_{exp}$}  & \colhead{$f_{\rm FOV}$} & \colhead{$L_{min}/10^{37}$} & \colhead{Ref.} \\
\colhead{} &  \colhead{} & \colhead{(Mpc)} & \colhead{(erg s$^{-1}$)}  & \colhead{(erg s$^{-1}$)} & \colhead{} & \colhead{} & \colhead{} & \colhead{(ks)} & \colhead{} & \colhead{(erg s$^{-1}$)} & \colhead{} } 
\startdata
 NGC 0253 &        SAB(s)c &   2.6 t &   7.64 &   4.55 &   5.2 &  969 & 1999-12-16 &  14.0 & 0.59      &    2.8 & 1\\
 NGC 0278 &       SAB(rs)b &  11.8 v &   5.82 &  21.97 & 138.8 & 2055 & 2001-06-22 &  38.3 & 1.00      &   28.6 & \\
 NGC 0404 &        SA(s)0- &   3.0 T &   0.30 &   0.14 &  83.6 &  870 & 1999-12-19 &  23.9 & 1.00      &    2.5 & 2\\
 NGC 0598 &        SA(s)cd &   0.9 C &   2.81 &   2.55 &   4.1 &  786 & 2000-08-30 &  46.3 & 0.18      &    0.1 & 3\\
 NGC 0628 &         SA(s)c &   9.7 v &   9.10 &  17.04 &  27.7 & 2058 & 2001-10-19 &  46.2 & 0.93      &   13.1 & \\
 NGC 0720 &             E5 &  24.6 I &  16.57 &   0.07 &   5.9 &  492 & 2000-10-12 &  39.6 & 1.00      &   88.6 & 4\\
 NGC 0891 &         SA(s)b &  10.0 P &  13.85 &  53.46 &  15.4 &  794 & 2000-11-01 &  50.9 & 0.82      &   13.6 & 5\\
 NGC 1068 &     (R)SA(rs)b &  14.4 v &  26.19 & 221.17 &   4.3 &  344 & 2000-02-21 &  47.4 & 1.00      &   27.0 & 6\\
\enddata
\end{deluxetable}
\clearpage
\begin{deluxetable}{llrrrrrrrlrl}
\tablecolumns{11}
\tablewidth{0pt}
\tabletypesize{\scriptsize}
\tablenum{1} 
\tablecaption{{\sl Continued}}
\tablehead{ \colhead{Galaxy} & \colhead{Type} &  \colhead{Distance$^a$} & \colhead{$L_{\rm B}/10^{42}$} & \colhead{$L_{\rm FIR}/10^{42}$} & \colhead{$\theta_p^b$}  & \colhead{ObsID} & \colhead{Date} & \colhead{$t_{exp}$}  & \colhead{$f_{\rm FOV}$} & \colhead{$L_{min}/10^{37}$} & \colhead{Ref.} \\
\colhead{} &  \colhead{} & \colhead{(Mpc)} & \colhead{(erg s$^{-1}$)}  & \colhead{(erg s$^{-1}$)} & \colhead{} & \colhead{} & \colhead{} & \colhead{(ks)} & \colhead{} & \colhead{(erg s$^{-1}$)} & \colhead{} } 
\startdata
 NGC 4314 &        SB(rs)a &  13.1 t &   4.53 &   4.39 &   9.3 & 2062 & 2001-04-02 &  16.1 & 1.00      &   31.3 & \\
          &                &         &        &        &       & 2063 & 2001-06-22 &  16.1 & 1.00      &        & \\
 NGC 4365 &             E3 &  20.9 F &  21.56 &   0.61 &  12.8 & 2015 & 2001-06-02 &  40.4 & 1.00      &   62.5 & 27\\
 NGC 4374 &             E1 &  17.4 F &  23.25 &   1.07 &  13.2 &  803 & 2000-05-19 &  28.5 & 1.00      &   63.8 & 28\\
 NGC 4382 &   SA(s)0+(pec) &  16.6 F &  21.56 &   0.24 &   1.1 & 2016 & 2001-05-29 &  39.7 & 1.00      &   41.5 & 29\\
 NGC 4388 &         SA(s)b &  16.7 v &  10.44 &  20.11 &   6.7 & 1619 & 2001-06-08 &  20.0 & 1.00      &   83.8 & 30\\
 NGC 4395 &         SA(s)m &   4.2 v &   0.81 &   0.64 &  10.6 &  882 & 2000-06-20 &  16.7 & 0.31(1/2) &    6.0 & \\
 NGC 4449 &            IBm &   3.7 t &   1.12 &   3.19 &  23.7 & 2031 & 2001-02-04 &  26.6 & 1.00      &    2.0 & 31\\
\enddata

\tablenotetext{a}{C:Cepheid period-luminosity; I:I-band SBF; T:Tip of RGB; P:Pla
netary nebula luminosity function; G:Globular cluster luminosity function; t:ter
tiary estimators; v:Tully-Fisher and variants including Tully (1988); F:average
of primary estimators from Ferrerase et~al. (2000)}
\tablenotetext{b}{Nearest-neighbor distance in units of the $D_{25}$ diameter}
\tablerefs{
(1)~Strickland et~al. 2000;
(2)~Eracleous et~al. 2002;
(3)~McDowell et~al. 2000;
(4)~Jeltema et~al. 2003;
(5)~Bregman \& Irwin 2002;
(6)~Smith \& Wilson 2003;
(7)~Irwin, Sarazin \& Bregman 2002;
(8)~Kim \& Fabbiano 2003;
(9)~Irwin, Athey \& Bregman 2003;
(10)~Blanton, Sarazin \& Irwin 2001;
(11)~Heike et~al. 2003;
(12)~Schlegel \& Pannuti 2003;
(13)~Humphrey et~al. 2003;
(14)~Swartz et~al. 2003;
(15)~Matsumoto et~al. 2001;
(16)~Cecil, Bland-Hawthorn \& Veilleux 2002;
(17)~Ott, Martin \& Walter  2003;
(18)~Schlegel 2001;
(19)~Wang, Chaves \& Irwin  2003;
(20)~Strickland et~al. 2001;
(21)~Fabbiano, Zezas \& Murray 2001;
(22)~Collinge et~al.  2001;
(23)~Yang, Wilson \& Ferruit  2001;
(24)~Cagnoni et~al. 2003;
(25)~Wilson, Yang \& Cecil  2001;
(26)~Jimez-Bailn et~al. 2003;
(27)~Kundu et~al. 2003;
(28)~Finoguenov \& Jones 2002;
(29)~Sivakoff, Sarazin \& Irwin 2003;
(30)~Iwasawa et~al.  2003;
(31)~Summers et~al. 2003;
(32)~Maccarone, Kundu \& Zepf 2003;
(33)~Roberts et~al. 2002;
(34)~Wilson \& Yang  2002;
(35)~Di~Stefano et~al.  2003;
(36)~Wang et~al.  2001;
(37)~Jones et~al.  2002;
(38)~Sarazin, Irwin \& Bregman  2001;
(39)~Pellegrini et~al.  2002;
(40)~Schurch, Roberts \& Warwick  2002;
(41)~Kraft et~al.  2001;
(42)~Roberts et~al. 2001;
(43)~Liu et~al. 2002;
(44)~Soria \& Wu  2002;
(45)~Mukai et~al.  2003;
(46)~Smith \& Wilson   2001;
(47)~Holt et~al.  2003;
(48)~Trinchieri et~al. 2003;
(49)~Fabbiano et~al.  2003
}

\end{deluxetable}

\begin{deluxetable}{lrrrrrrrrrr}
\tablecolumns{11}
\tablewidth{0pt}
\tabletypesize{\scriptsize}
\tablenum{2} 
\tablecaption{Properties of ULX Candidates} 
\tablehead{
\colhead{Galaxy} & \colhead{R.A.} &  \colhead{Decl.} & \colhead{Spectral$^a$} & \colhead{$n_H$} & \colhead{$\chi^2/{\rm dof}$}  & \colhead{$L_{{\rm obs}}$} & \colhead{$L_{{\rm int}}$} & \colhead{$(H-M)$}  & \colhead{$(M-S)$} & \colhead{$P_{KS}$} \\
\colhead{} &  \colhead{(h m s)} & \colhead{(d m s)} & \colhead{Parameter} & \colhead{($10^{22}$ cm$^{-2}$)}  & \colhead{}& \colhead{($10^{39}$ erg s$^{-1}$)} & \colhead{($10^{39}$ erg s$^{-1}$)} & \colhead{color} & \colhead{color} & \colhead{} }
\startdata
 NGC 628 &  1 36 51.06 & +15 45 46.8 & $     2.01^{+    0.12}_{   -0.12} $ & $     0.05^{+    0.00}_{   -0.00} $ &    41.06/35 & $     1.50 \pm     0.08 $ & $     1.78 \pm     0.08 $   &    -0.17 &    -0.12 &     0.001 \\
 NGC 720 &  1 52 55.86 & -13 43 51.1 & $     1.56^{+    0.50}_{   -0.42} $ & $     0.01^{+    0.00}_{   -0.00} $ &     3.19/5 & $     0.99 \pm     0.22 $ & $     1.08 \pm     0.24 $   &    -0.05 &    -0.14 &     0.521 \\
 NGC 720 &  1 52 56.54 & -13 43 47.8 & $     1.78^{+    1.24}_{   -0.82} $ & $     0.27^{+    0.75}_{   -0.27} $ &     4.38/3 & $     1.16 \pm     0.24 $ & $     1.57 \pm     0.52 $   &    -0.13 &     0.28 &     0.712 \\
 NGC 720 &  1 52 59.44 & -13 43 57.5 & $     2.33^{+    0.98}_{   -0.71} $ & $     0.14^{+    0.23}_{   -0.14} $ &     2.04/2 & $     1.07 \pm     0.19 $ & $     1.46 \pm     0.34 $   &    -0.25 &    -0.09 &     0.637 \\
 NGC 720 &  1 53 06.48 & -13 45 40.9 & $     1.79^{+    0.92}_{   -0.60} $ & $     0.13^{+    0.28}_{   -0.13} $ &     3.55/4 & $     1.12 \pm     0.23 $ & $     1.32 \pm     0.22 $   &    -0.17 &     0.16 &     0.274 \\
 NGC 891 &  2 22 31.25 & +42 19 57.8 & $     3.55^{+    0.41\, c}_{   -0.55} $ & $     0.61^{+    0.05}_{   -0.03} $ &    62.21/60 & $     4.05 \pm     0.26 $ & $     5.96 \pm     0.25 $   &    -0.22 &     0.40 &     0.688 \\
 NGC 891 &  2 22 31.35 & +42 20 24.4 & $     1.93^{+    0.14}_{   -0.13} $ & $     0.69^{+    0.08}_{   -0.07} $ &    79.88/77 & $     5.03 \pm     0.16 $ & $     8.33 \pm     0.35 $   &    -0.08 &     0.42 &     0.603 \\
 NGC 1068 &  2 42 38.89 & -00 00 55.1 & $     0.97^{+    0.15}_{   -0.14} $ & $     0.63^{+    0.13}_{   -0.10} $ &    75.10/61 & $    11.33 \pm     0.42 $ & $    13.51 \pm     0.47 $   &     0.28 &     0.31 &     0.822 \\
\enddata
 \end{deluxetable}
\clearpage
\begin{deluxetable}{lrrrrrrrrrr}
\tablecolumns{11}
\tablewidth{0pt}
\tabletypesize{\scriptsize}
\tablenum{2} 
\tablecaption{{\sl Continued}} 
\tablehead{
\colhead{Galaxy} & \colhead{R.A.} &  \colhead{Decl.} & \colhead{Spectral$^a$} & \colhead{$n_H$} & \colhead{$\chi^2/{\rm dof}$}  & \colhead{$L_{{\rm obs}}$} & \colhead{$L_{{\rm int}}$} & \colhead{$(H-M)$}  & \colhead{$(M-S)$} & \colhead{$P_{KS}$} \\
\colhead{} &  \colhead{(h m s)} & \colhead{(d m s)} & \colhead{Parameter} & \colhead{($10^{22}$ cm$^{-2}$)}  & \colhead{}& \colhead{($10^{39}$ erg s$^{-1}$)} & \colhead{($10^{39}$ erg s$^{-1}$)} & \colhead{color} & \colhead{color} & \colhead{} }
\startdata
 NGC 4038 & 12 01 53.50 & -18 53 10.7 & $     4.67^{+    1.21\, c}_{   -1.90} $ & $     0.06^{+    0.04}_{   -0.03} $ &    55.86/51 & $     1.89 \pm     0.16 $ & $     2.07 \pm     0.10 $   &    -0.25 &    -0.13 &     0.250 \\
 NGC 4038 & 12 01 54.27 & -18 52 01.8 & $     1.99^{+    0.23}_{   -0.21} $ & $     0.09^{+    0.06}_{   -0.06} $ &    33.96/27 & $     1.26 \pm     0.09 $ & $     1.53 \pm     0.10 $   &    -0.20 &     0.00 &     0.987 \\
 NGC 4038 & 12 01 54.35 & -18 52 10.2 & $     1.18^{+    0.21}_{   -0.19} $ & $     0.07^{+    0.07}_{   -0.07} $ &    26.56/25 & $     1.98 \pm     0.15 $ & $     2.13 \pm     0.13 $   &    -0.02 &     0.13 &     0.849 \\
 NGC 4038 & 12 01 54.77 & -18 52 52.1 & $     2.54^{+    0.75}_{   -0.65} $ & $     6.51^{+    2.12}_{   -1.67} $ &     9.15/13 & $     2.93 \pm     0.69 $ & $    17.86 \pm    10.75 $   &     0.87 &     0.07 &     0.517 \\
 NGC 4038 & 12 01 54.97 & -18 53 14.9 & $     1.69^{+    0.12}_{   -0.12} $ & $     0.21^{+    0.04}_{   -0.04} $ &    84.34/77 & $     5.42 \pm     0.21 $ & $     6.75 \pm     0.19 $   &    -0.17 &     0.24 &     0.003 \\
 NGC 4038 & 12 01 55.65 & -18 52 15.0 & $     1.26^{+    0.11}_{   -0.10} $ & $     0.03^{+    0.03}_{   -0.03} $ &    74.89/84 & $     7.26 \pm     0.29 $ & $     7.64 \pm     0.23 $   &    -0.08 &     0.06 &     0.807 \\
 NGC 4038 & 12 01 56.42 & -18 51 57.8 & $     1.95^{+    0.11}_{   -0.11} $ & $     0.11^{+    0.03}_{   -0.03} $ &    95.67/85 & $     5.09 \pm     0.19 $ & $     6.26 \pm     0.20 $   &    -0.21 &     0.03 &     0.124 \\
 NGC 4039 & 12 01 54.48 & -18 53 05.4 & $     3.23^{+    1.34}_{   -0.67} $ & $     0.19^{+    0.20}_{   -0.11} $ &    57.82/14 & $     0.40 \pm     0.04 $ & $     0.85 \pm     0.29 $   &    -0.28 &    -0.19 &     0.187 \\
\enddata
 \end{deluxetable}
\clearpage
\begin{deluxetable}{lrrrrrrrrrr}
\tablecolumns{11}
\tablewidth{0pt}
\tabletypesize{\scriptsize}
\tablenum{2} 
\tablecaption{{\sl Continued}} 
\tablehead{
\colhead{Galaxy} & \colhead{R.A.} &  \colhead{Decl.} & \colhead{Spectral$^a$} & \colhead{$n_H$} & \colhead{$\chi^2/{\rm dof}$}  & \colhead{$L_{{\rm obs}}$} & \colhead{$L_{{\rm int}}$} & \colhead{$(H-M)$}  & \colhead{$(M-S)$} & \colhead{$P_{KS}$} \\
\colhead{} &  \colhead{(h m s)} & \colhead{(d m s)} & \colhead{Parameter} & \colhead{($10^{22}$ cm$^{-2}$)}  & \colhead{}& \colhead{($10^{39}$ erg s$^{-1}$)} & \colhead{($10^{39}$ erg s$^{-1}$)} & \colhead{color} & \colhead{color} & \colhead{} }
\startdata
 NGC 4594 & 12 40 00.97 & -11 36 54.2 & $     2.52^{+    0.47}_{   -0.41} $ & $     0.18^{+    0.13}_{   -0.11} $ &     8.43/9 & $     0.91 \pm     0.08 $ & $     1.44 \pm     0.23 $   &    -0.30 &     0.05 &     0.352 \\
 NGC 4594 & 12 40 01.10 & -11 37 24.2 & $     5.62^{+    1.49}_{   -1.13} $ & $     0.36^{+    0.22}_{   -0.17} $ &     7.55/12 & $     0.66 \pm     0.13 $ & $     5.73 \pm     4.64^d $   &    -0.22 &    -0.53 &     0.096 \\
 NGC 4621 & 12 42 02.99 & +11 41 16.3 & $     1.59^{+    0.71}_{   -0.58} $ & $     0.21^{+    0.24}_{   -0.19} $ &     4.93/5 & $     1.00 \pm     0.18 $ & $     1.24 \pm     0.17 $   &    -0.03 &     0.20 &     0.938 \\
 NGC 4631 & 12 41 55.56 & +32 32 16.9 & $     1.75^{+    0.11}_{   -0.10} $ & $     0.16^{+    0.03}_{   -0.03} $ &   134.36/105 & $     2.42 \pm     0.10 $ & $     2.94 \pm     0.08 $   &    -0.18 &     0.14 &     0.599 \\
 NGC 4631 & 12 42 11.13 & +32 32 35.8 & $     2.79^{+    0.38}_{   -0.35} $ & $     3.99^{+    0.63}_{   -0.56} $ &    44.86/47 & $     2.20 \pm     0.12 $ & $    13.81 \pm     3.96 $   &     0.57 &     0.20 &     0.228 \\
 NGC 4649 & 12 43 32.08 & +11 34 18.5 & $     2.33^{+    0.76}_{   -0.62} $ & $     0.67^{+    0.38}_{   -0.27} $ &     3.12/6 & $     0.86 \pm     0.14 $ & $     1.73 \pm     0.47 $   &    -0.16 &     0.40 &     0.164 \\
 NGC 4649 & 12 43 37.27 & +11 31 43.7 & $     1.85^{+    0.25}_{   -0.22} $ & $     0.02^{+    0.00}_{   -0.00} $ &     4.91/7 & $     0.97 \pm     0.11 $ & $     1.04 \pm     0.12 $   &    -0.15 &    -0.26 &     0.016 \\
 NGC 4649 & 12 43 45.00 & +11 32 33.4 & $     1.53^{+    0.28}_{   -0.26} $ & $     0.02^{+    0.00}_{   -0.00} $ &    14.13/11 & $     0.90 \pm     0.11 $ & $     0.93 \pm     0.14 $   &    -0.10 &    -0.10 &     0.986 \\
\enddata

\tablenotetext{a}{Power-law index, $\Gamma$, unless otherwise noted.}
\tablenotetext{b}{Disk-blackbody model inner disk temperature, $kT_{in}$, in keV
.}
\tablenotetext{c}{Thermal emission-line model temperature, $kT$, in keV.}
\tablenotetext{d}{Intrinsic luminosity $<$$10^{39}$~\ergl\ when fit using a
thermal emission-line model.}

\end{deluxetable}

\clearpage
\begin{center}
\small{
\begin{tabular}{lcccccc}
\multicolumn{7}{c}{{\sc Table 3}} \\
\multicolumn{7}{c}{{\sc Linear Correlation Coefficients$^a$}} \\
\hline \hline
 & \multicolumn{3}{c}{All Galaxies} & \multicolumn{3}{c}{Subsample} \\
 & All & Ellip. & Spiral & All & Ellip. & Spiral \\
\hline
$N_{\rm ULX}$,\lb\               & 0.74 & 0.89 & ---  & 0.70   & (0.89) & 0.71 \\
$N_{\rm ULX}$,\lfir\             & 0.63 & ---  & 0.88 & (0.50) & ---  & 0.71 \\
$N_{\rm ULX}$,$\theta_p$         & ---  & ---  &(-.54)& ---    & ---  & --- \\
$L_{\rm ULX}$,\lb\               & ---  & 0.82 & ---  & ---    & ---  & 0.76 \\
$L_{\rm ULX}$,\lfir\             & 0.57 & ---  & 0.88 & 0.83   & ---  & 0.80 \\
$L_{\rm ULX}$,$\theta_p$         & ---  & ---  &(-.54)& ---    & ---  & --- \\
$L_{\rm ULX}/N_{\rm ULX}$,\lb\   & ---  & 0.86 & ---  & ---    & ---  & 0.82 \\
$L_{\rm ULX}/N_{\rm ULX}$,\lfir\ & ---  & ---  & 0.82 & 0.92   & ---  & 0.91 \\
\hline
$L_{\rm ULX}/$\lb,\lfir$/$\lb\ & 0.80 & --- & 0.99 & --- & --- & --- \\
$N_{\rm ULX}/$\lb,\lfir$/$\lb\ & 0.56 & --- & 0.97 & --- & --- & --- \\
$(L_{\rm ULX}/N_{\rm ULX})/$\lb,\lfir$/$\lb\ & --- & --- & 0.90 & --- & --- & --- \\
\hline
\multicolumn{7}{l}{$^a$() denote marginally significant correlation,
--- denotes no correlation} \\
\end{tabular}
} 
\end{center}

\end{document}